\documentclass[preprint]{elsarticle}
\usepackage{footmisc}
\makeatletter
\let\@fnsymbol\@alph
\makeatother
\usepackage[english]{babel}
\usepackage[utf8]{inputenc}
\usepackage{fancyhdr}
\usepackage{amsmath}
\usepackage{amssymb}
\usepackage{upgreek}
\usepackage{graphicx}
\usepackage{endnotes}
\usepackage{float}
\usepackage{caption}
\usepackage{subcaption}
\usepackage{color}
\usepackage{url}
\usepackage{lineno}
\usepackage{appendix}
\usepackage[colorlinks]{hyperref}
\usepackage{xspace}
\usepackage{mathtools}
\AtBeginDocument{%
  \hypersetup{%
    linkcolor=red,%
    citecolor=green,%
    urlcolor=cyan,%
  }%
}%

\usepackage{geometry}
\def\costheta{\ensuremath{\cos \theta}\xspace}
\def\costhetaq{\ensuremath{\cos \theta_{q}}\xspace}

\def\costhetac{\ensuremath{\cos \theta_{c}}\xspace}

\def\fb{fb\ensuremath{^{-1}}\xspace}

\def\bbbar{\ensuremath{b}\ensuremath{\overline{b}}\xspace}
\def\qqbar{\ensuremath{q}\ensuremath{\overline{q}}\xspace}
\def\ccbar{\ensuremath{c}\ensuremath{\overline{c}}\xspace}
\def\eebb{\ensuremath{e^{\mbox{\scriptsize +}}}\ensuremath{e^{\mbox{\scriptsize -}}}\ensuremath{\rightarrow}\ensuremath{b}\ensuremath{\overline{b}}\xspace}
\def\eecc{\ensuremath{e^{\mbox{\scriptsize +}}}\ensuremath{e^{\mbox{\scriptsize -}}}\ensuremath{\rightarrow}\ensuremath{c}\ensuremath{\overline{c}}\xspace}
\def\eeqq{\ensuremath{e^{\mbox{\scriptsize +}}}\ensuremath{e^{\mbox{\scriptsize -}}}\ensuremath{\rightarrow}\ensuremath{q}\ensuremath{\overline{q}}\xspace}

\def\eeZqq{\ensuremath{e^{\mbox{\scriptsize +}}}\ensuremath{e^{\mbox{\scriptsize -}}}\ensuremath{\rightarrow}\ensuremath{Z}\ensuremath{\rightarrow}\ensuremath{q}\ensuremath{\overline{q}}\xspace}

\def\eLpR{\ensuremath{e_L}\ensuremath{p_R}\xspace}
\def\eRpL{\ensuremath{e_R}\ensuremath{p_L}\xspace}

\def\dedx{\ensuremath{dE/\/dx}\xspace}
\def\Afbb{\ensuremath{A^{b\bar{b}}_{FB}}\xspace}

\def\Rb{\ensuremath{R_{b}}\xspace}
\def\Rq{\ensuremath{R_{q}}\xspace}
\def\Afbc{\ensuremath{A^{c\bar{c}}_{FB}}\xspace}
\def\Afbq{\ensuremath{A^{q\bar{q}}_{FB}}\xspace}
\def\Rc{\ensuremath{R_{c}}\xspace}

\usepackage{fancyhdr}
\fancypagestyle{pprintTitle}{%
  \lhead{Contribution for LCWS2019, Tohoku (Japan)} \chead{}\rhead{ILD-PHYS-PROC-2020-001}
\lfoot{}\cfoot{}\rfoot{}

}

\author[a,b]{A. Irles\footnote{Corresponding author.}}
\author[a,b]{R. P\"oschl}
\author[a,b]{F. Richard}
\address[a]{{\bf on behalf of the ILD concept group}}
\address[b]{Universit{\'e} Paris-Saclay, CNRS/IN2P3, IJCLab, 91405 Orsay, France}

\title{\LARGE\bf Production and measurement of \eecc signatures at the 250 GeV ILC}

\begin{document}

\begin{frontmatter}
  



\begin{abstract}
 This document discusses for the first time the experimental prospects on the measurement of cross section
 and the forward-backward asymmetry in \eecc 
 collisions at 250 GeV at the International Linear Collider operating polarised beams.
 The cross section will be normalised to the total hadronic cross section.
 We discuss the results for an analysis assuming the integrated luminosity of 2000 \fb foreseen in the baseline project.
 The measurement requires determining the charge of both jets identified as originated by a $c$-quark. 
 The charge measurement is optimally performed using the precise micro-vertex 
 detector of the detector ILD and the charged kaon identification provided by the \dedx information of its TPC. 
 Thanks to the beam polarisation, we can separate  the four independent chirality combinations of the electroweak
 couplings, enhancing in this way the sensitivity to new physics effects.
 We show that due to the unprecedented precision that will be achieved by the ILC for these observables,
 the ILC will be sensitive to the existence of beyond the standard model Randal Sundrum resonances
 of several tens of TeV. 
\end{abstract}

\end{frontmatter}



\section{Introduction}
\label{sec:intro}

The $c$-quark (and $b$-quark) electroweak couplings to the $Z$-boson have been determined \cite{ALEPH:2005ab} by
the LEP1
detector collaborations and the SLD Collaboration in \eecc (and \eebb) collisions at the $Z$-pole.
These couplings are usually determined from the measurement of
experimental distributions such as the \eecc (and \eebb) cross section divided by the
total hadronic cross section and the forward-backward asymmetry.
The measurements done by LEP1 detector collaborations
profited from higher luminosities recorded than the SLD ($\sim\times$20 times more).
Despite the large difference
on integrated recorded luminosity, the SLD obtained a similar precision measurements due to the
benefits of having a polarised beam and a smaller radius of the vacuum beam pipe that permitted
instrumenting the tracker closer to the interaction point.

The cross section normalised to the total hadronic cross section is defined as:
\begin{equation}
  \Rq= \frac{\sigma^{q\bar{q}}}{\sigma^{had.}}
\end{equation}
where $\sigma^{q\bar{q}}$ is the \eeqq total cross section for given $q$-quark flavour and
$\sigma^{had.}$ is the cross section for all $q$-quark flavours. 

The forward-backward asymmetry is defined as:
\begin{equation}
  \Afbq= \frac{\sigma^{q\bar{q}}_{F}-\sigma^{q\bar{q}}_{B}}{\sigma^{q\bar{q}}_{F}+\sigma^{q\bar{q}}_{B}}
\end{equation}
where $\sigma^{q\bar{q}}_{F,B}$ is the differential \eeqq cross section integrated over the forward/backward
hemisphere as defined by the polar scattering angle $\theta$ of the $\vec{p}_{b\bar{b}} = \vec{p}_{b}-\vec{p}_{\bar{b}}$.
The $z$-axis of the right-handed coordinate frame is given by the direction of the incoming electron beam.

These two observables have been measured at the $Z$-pole by the LEP experiments and by the SLD.
At the $Z$-pole, the \Rq observable is interpreted as $\Rq= \frac{\Gamma_{q\bar{q}}}{\Gamma_{had.}}$
where $\Gamma_{q\bar{q}}$ is the decay width of $Z\rightarrow\qqbar$ and $\Gamma_{had.}$ is the total hadronic $Z$-decay width.
The \Rc and \Afbc observables were measured with experimental precisions of $2\%$ and $4\%$ respectively.

The International Linear Collider (ILC) \cite{Behnke:2013xla,Baer:2013cma,Adolphsen:2013jya,Adolphsen:2013kya,Behnke:2013lya} 
is a linear electron-positron collider 
with polarised beams that will produce collisions at several energies.
In this document, we discuss the prospects for the measurements of these two observables
an the ILC operating at a centre of mass energy of 250 GeV (ILC250).
At energies far above the $Z$-pole the experimental observables
are sensitive to the interference between the $\gamma$ the Z and potential new vector bosons.

The International Large Detector (ILD) \cite{Behnke:2013lya} is one of the proposed detectors
to measure the interactions. This detector will be optimised
to use Particle Flow (PF) reconstruction algorithms \cite{Brient:2002gh,Morgunov:2004ed,Sefkow:2015hna} in order to
reconstruct and separate individual particles produced in the collisions. For this,
a high granularity calorimetric system is foreseen to be placed inside
a $\sim$4 T magnetic field.
Moreover, the ILD will have a high precision tracking system with the first layer placed at 16 mm 
from the interaction point to 
maximise the tracking and vertex reconstruction capabilities. 
The central tracker of the ILD is a Time Projection Chamber (TPC)
that provides pattern recognition with more than 200 space points.

\section{Event reconstruction and selection}
\label{sec:reco}

All results shown here are obtained using detailed simulation of the 
ILD concept \cite{Behnke:2013lya}.
The experimental studies are made for the case in which the electron and positron beams
are 100\% polarised. We use the following notation:
\eLpR for the cases in which the electron beam has 100\% left polarisation and
the positron beam has 100\% right polarisation (and vice versa for \eRpL).
The size of the analysed samples is the equivalent of 250 \fb for each of the processes
while the ILC250 programme foresees a total of 2000 \fb shared between the different
beam polarisations schemes: 900 \fb for each of the $P(e^{-},e^{+}) = (\pm80\%, \mp30\%)$
and 100 \fb for $P(e^{-},e^{+}) = (\pm80\%, \pm30\%)$.
Final results will be scaled to the foreseen luminosity.

The events are generated at leading order using the
WHIZARD 1.95 \cite{Kilian:2007gr,Moretti:2001zz} event generator. The
parton showering and hadronisation
are simulated by the Pythia 6.422 event generator \cite{Sj_strand_2006}.
The ILD detector geometry and the interaction of the particles with the detector are simulated
within the Mokka framework interfaced with the Geant4 toolkit \cite{Agostinelli:2002hh,Allison:2006ve,Allison:2016lfl}. 
The different reconstruction algorithms are implemented in the ILCSoft toolkit. 
We make use of the tracking, quark-tagging, particle identification in the TPC and jet 
clustering algorithms described in Ref.\cite{bilokin:tel-01826535}.
The primary and secondary vertex reconstruction has special importance for this analysis.
Therefore they were optimised to fit the high precision requirements of this analysis (see, again, Ref. \cite{bilokin:tel-01826535} for more details).

The \eeqq events have a very distinguishable signature, in which both quarks
are observed as a two jet back-to-back system at same energy.
Events are reconstructed using the Durham jet algorithm forced to form two jets.
Leptonic events are removed by a selection of two hadronic jets.
Backgrounds are given by: the events that are subject to a radiative return to the $Z$-pole
due to initial state radiation and di-boson events with hadronic decays. These backgrounds are rejected 
by a combination of three cuts:

\begin{enumerate}
  \item A cut on the $y_{23}$ distance that defines the Durham distance at which a two jet system would be reconstructed as
    a three jet system;
  \item A cut on the sum of the two jet masses. This cut helps 
    reducing the impact of QCD final state radiation that dilutes the back-to-back configuration of the two jets
    and also helps suppressing the background from $WW/ZH/ZZ$ events.
  \item A cut on the sphericity of the event. This cut depends on the polar angle at which the jets are
    reconstructed since by ISR the system receives a boost, {\it i.e.}
    is not back-to-back anymore and/or receives a transverse momentum.
\end{enumerate}
   
\begin{figure}[!h]
  \centering
      \begin{tabular}{cc}
        \includegraphics[width=0.45\textwidth]{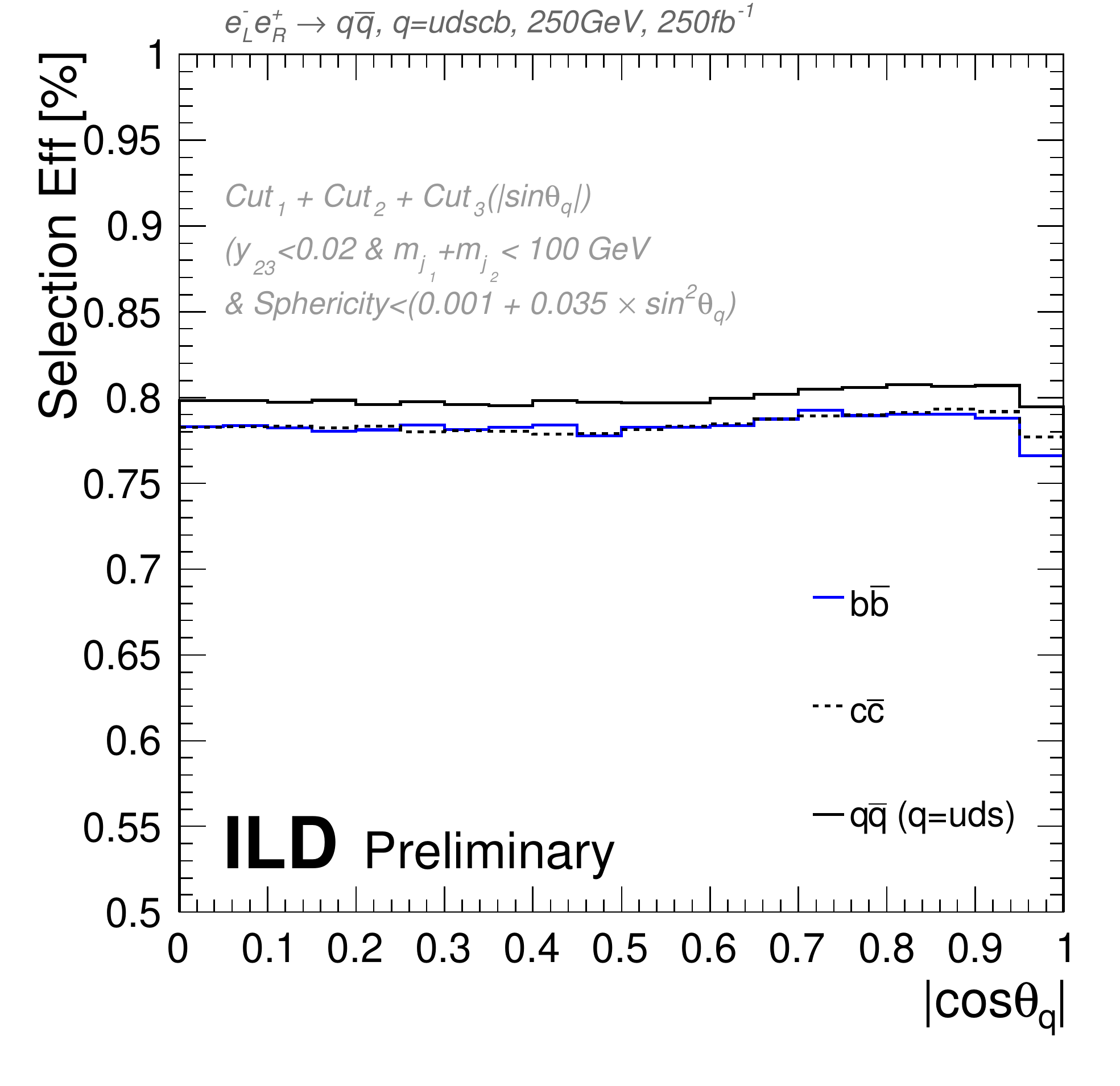} & \includegraphics[width=0.45\textwidth]{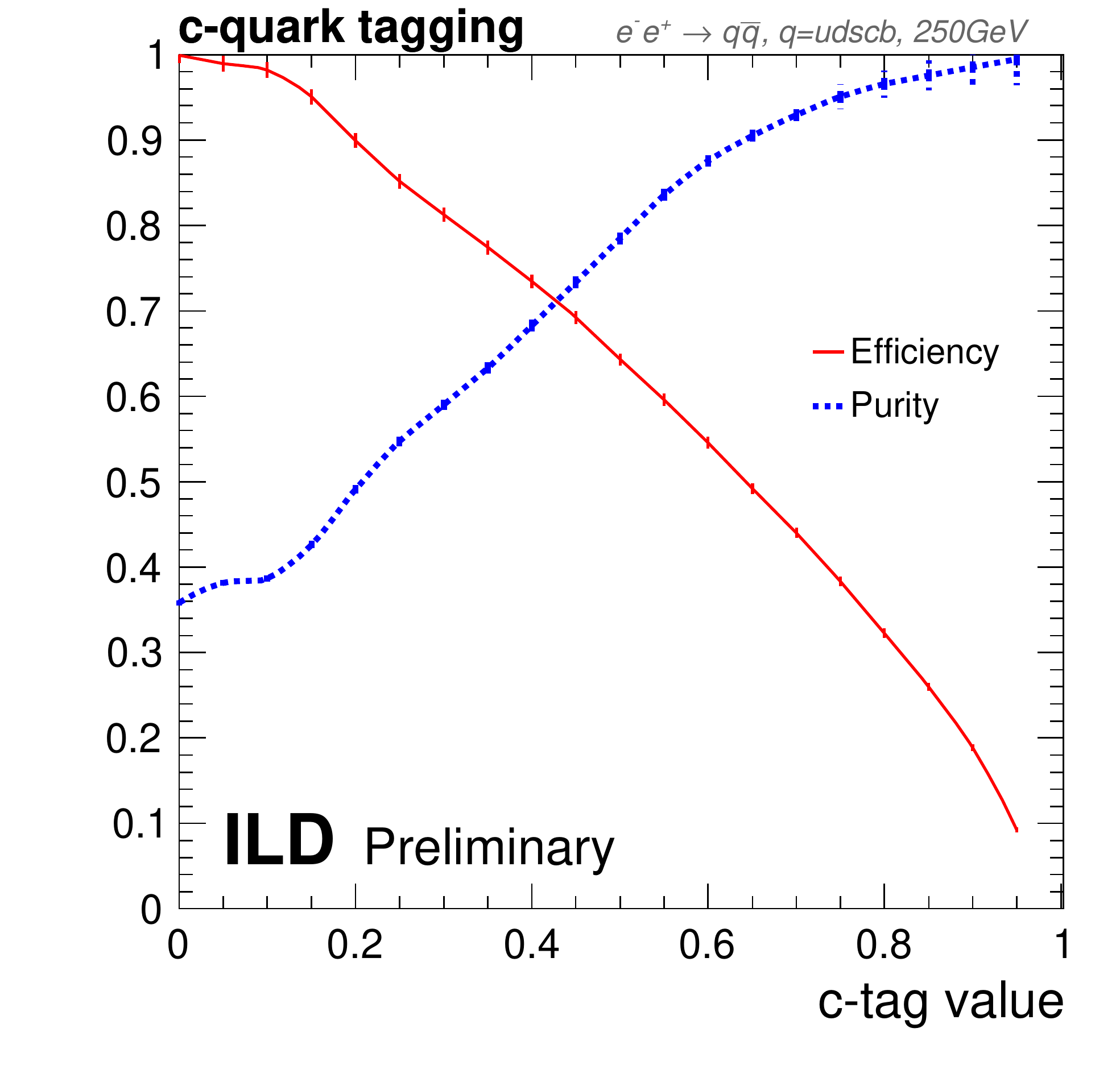}
      \end{tabular}
      \caption{\label{fig:selection} Left plot: efficiency of the preselection for the different quark flavours vs the angular distribution of the two jet system. Right plot: efficiency and purity of the $c$-quark tagging.}
\end{figure}

The values of these cuts can be found in the legend of the Figure \ref{fig:selection}, left.
A more detailed discussion on the selection procedure can be found at the Appendix \ref{app:sel}.
With these cuts, we select a clean sample of \eeqq with $q=u,d,s,c,b$.
The efficiency of selection is shown as a function of the \costhetaq angle absolute value.
The selection efficiency is slightly different between the light and heavy quarks.
These differences come from the quark mass differences: the larger the quark mass, the harder and less collinear
the gluon final state radiation \cite{Abreu:1997ey,rodrigomb} and are well explained by QCD.

For the final step of our pre-selection of \eecc, we suppress the most dominant 
background, the \eeqq ($q=u,d,s,b$), by requiring that one or two jets have
a large $c$-tag value (c-tag$>0.875$). The quality, in terms of efficiency
and purity, of the $c$-quark tagging for 125 GeV jets as a function of the $c$-tag value is shown
in the right plot from Figure \ref{fig:selection}.

The impact of all the selection cuts on the signal selection efficiency and the ratio of 
background vs signal is summarised in Table \ref{tab:cutflow} and in Figure \ref{fig:selection2}.

\begin{table}[!ht]
  \scriptsize
  \begin{tabular}{c|c|cccccc}
    \hline
    \multicolumn{8}{c}{{\bf \eLpR}}\\
      \hline
      & {\bf Signal [\%]} & \multicolumn{6}{c}{{\bf Background processes [\%] (B/S [\%]) }}\\
      & {\ccbar} & {\bbbar}  &{\qqbar} ({\it uds}) & $\gamma\qqbar$ & ZZ & WW & HZ \\
      \hline
       $Cut_{1}$      
      & 85.6 & 85,2 (69,1) & 87,2 (224,0) & 69,0 (53,2) & 15,4 (3,1) & 16,7 (35,1) & 11,0 (0,2)\\
       $Cut_{1}+Cut_{2}$      
      & 82,1 & 82,1 (69,4) & 83,7 (224,4) & 65,7 (52,8) & 8,2 (1,7) & 10,5 (23,1) & 5,3 (0,1) \\
       $Cut_{1}+Cut_{2}+Cut_{3}$      
      &  78.5 & 78,4 (69,4) & 80,0 (224,4) & 24,5 (20,6) & 4,1 (0,9) & 3,2 (7,4) & 2,3 (0,0)  \\
       $Cut_{1}+Cut_{2}+Cut_{3}+ 1 ctag$ 
      & 38.9 & 2,1 (3,8) & 0,2 (0,9) & 1,1 (1,9) & 0,3 (0,1) & 0,2 (0,8)  & 0,1 (0,0) \\                 
       $Cut_{1}+Cut_{2}+Cut_{3}+ 2 ctag$ 
      & 7.3  & 0,0 (0,2) & 0,0 (0,0) & 0,1 (0,7) & 0,0 (0,0) & 0,0 (0,1) & 0,0 (0,0) \\
      \hline
      \multicolumn{8}{c}{}\\
      \hline
      \multicolumn{8}{c}{{\bf \eRpL}}\\
      \hline
      & {\bf Signal [\%]} & \multicolumn{6}{c}{{\bf Background processes [\%] (B/S [\%]) }}\\
      & {\ccbar} & {\bbbar}  &{\qqbar} ({\it uds}) & $\gamma\qqbar$ & ZZ & WW & HZ  \\
            \hline
       $Cut_{1}$      
      & 85.6 & 85,1 (35,8) & 87,1 (161,4) & 69,0 (38,8) & 17,7 (3,6) & 7,8 (0,4) & 10,9 (0,4)  \\
       $Cut_{1}+Cut_{2}$      
      & 82.2 & 82,0 (36,0) & 83,7 (161,5) & 65,7 (38,5) & 10,7 (2,3) & 4,7 (0,2) & 5,3 (0,2) \\
       $Cut_{1}+Cut_{2}+Cut_{3}$      
      & 78.6 & 78,2 (35,9) & 80,0 (161,5) & 24,4 (15,0) & 5,5 (1,2) & 1,4 (0,1) & 2,3 (0,1) \\
       $Cut_{1}+Cut_{2}+Cut_{3}+ 1 ctag$      
      & 38.9 & 2,1 (2,0) & 0,2 (0,6) & 1,3 (1,6) & 0,4 (0,2) & 0,2 (0,0)  & 0,1 (0,0) \\                 
       $Cut_{1}+Cut_{2}+Cut_{3}+ 2 ctag$ 
      & 7.3  & 0,0 (0,1) & 0,0 (0,0) & 0,1 (0,7) & 0,0 (0,1) & 0,0 (0,0) & 0,0 (0,0) \\
      \hline
    \end{tabular}
 \caption{\label{tab:cutflow} Percentage of the signal and the different background events remaining after each selection step. For completeness, the background to signal ratio (in $\%$ units) is also quoted. It is important to remark that the efficiencies shown for $\gamma\qqbar$ are calculated for
 a sample which has already a cut at the generation level to remove most of the radiative return events. All these events are removed with the first two cuts.}
\end{table}

\section{\Rc measurement}
\label{sec:Rc}

To reach the maximum of precision on the measurement of \Rc, we will measure it
at the same time as we measure the $c$-quark tagging efficiency
applying the Double Tag approach described in \cite{ALEPH:2005ab}.
The method is as follows: first we count the fraction
of jets in the preselected events that are tagged as a $c$-jet. This ratio is denominated $f_{1}$.
To use this method we have to assume that we know the value of \Rb, $\epsilon_{b}$ and $\epsilon_{uds}$
or that we measure them at the same time.
This measurement will also depend on the other sources of backgrounds which are very much
reduced by the preselection and that are expected to be known to the percent level or better, so they can be ignored 
at the first approach.
Secondly, we measure the fraction of preselected events in which both jets have been tagged as $c$-jets. This 
quantity is $f_{2}$ and has similar dependencies as $f_{1}$ with the addition
of the correlation variable $\rho_c$. This correlation factor accounts for the correlations
due to displacements of primary vertex determination (common for both jets),
purely geometric correlations associated to differences between the detector
inhomogeneities and QCD related effects associated to hard gluon emission.
The two ratios and their dependence on the different terms are described in the following equations:

\begin{equation}
  \begin{aligned}
f_{1}=\epsilon_{c}\Rc + \epsilon_{b} \Rb + \epsilon_{uds} (1-\Rc- \Rb) + F(\epsilon_{c}, \epsilon_{b}, \epsilon_{uds}, BKG)\\
f_{2}=\epsilon_{c}^{2}(1+\rho_c)\Rc + \epsilon_{b}^{2} \Rb + \epsilon_{uds}^{2} (1-\Rc- \Rb) + F(\epsilon_{c}^{2}, \epsilon_{b}^{2}, \epsilon_{uds}^{2}, BKG)
\label{eq:Rc}
  \end{aligned}
\end{equation}

In this method, the statistical uncertainty is determined by the size of the double tagged sample, which is
proportional to the square of the tagging efficiency. In the past, only the SLD Collaboration
was able to present a high precision \Rc measurement using the Double Tag technique \cite{Abe:2005nqa}, thanks to a superior
$c$-quark tagging. For the ILD case, the $\epsilon_{c}$ is of the order of the 35\% which is almost twice what the SLD achieved \cite{Abe:2005nqa}.

\begin{figure}[!pt]
\begin{center}
    \begin{tabular}{cc}
      \includegraphics[width=0.45\textwidth]{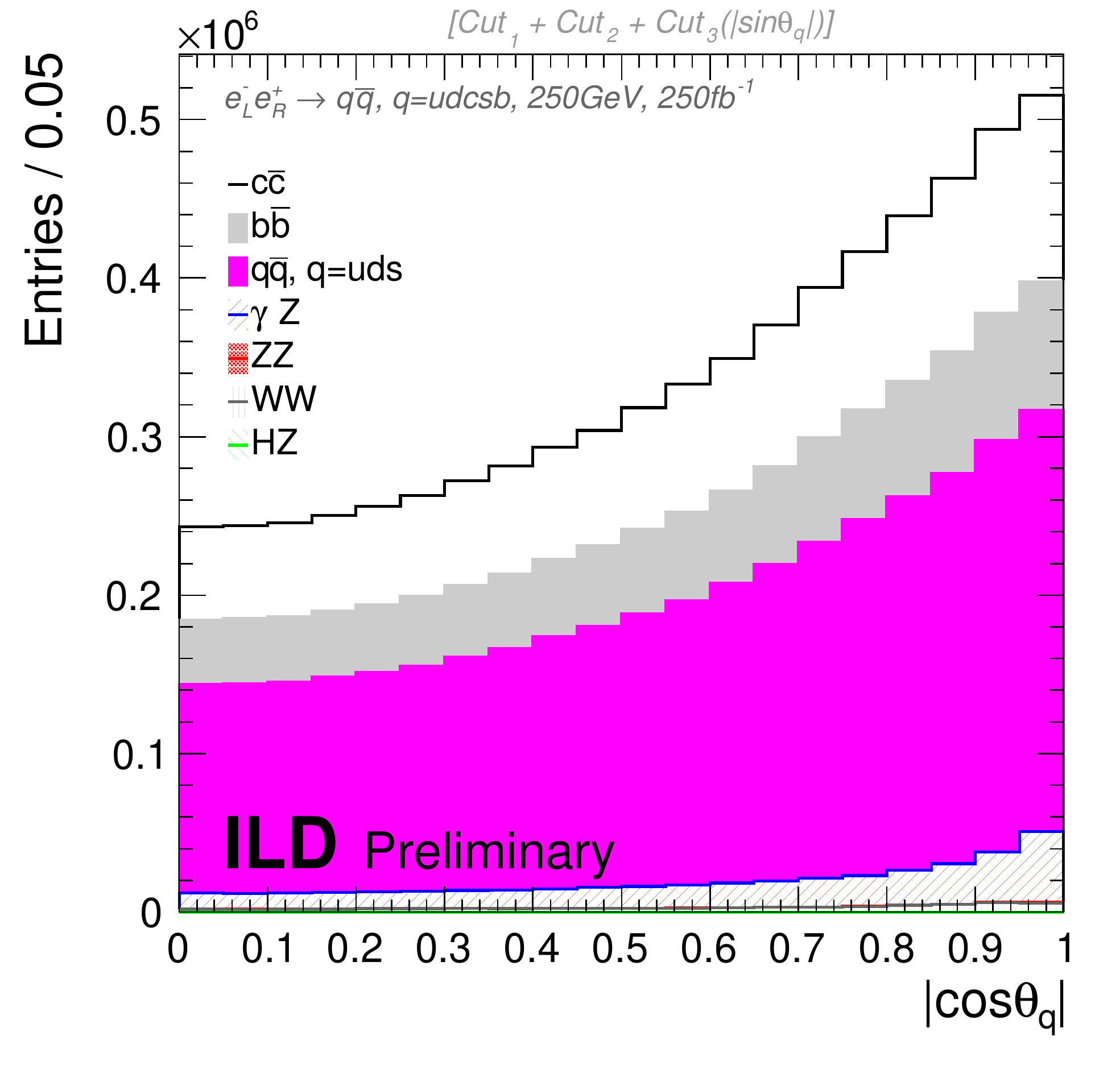} &
      \includegraphics[width=0.45\textwidth]{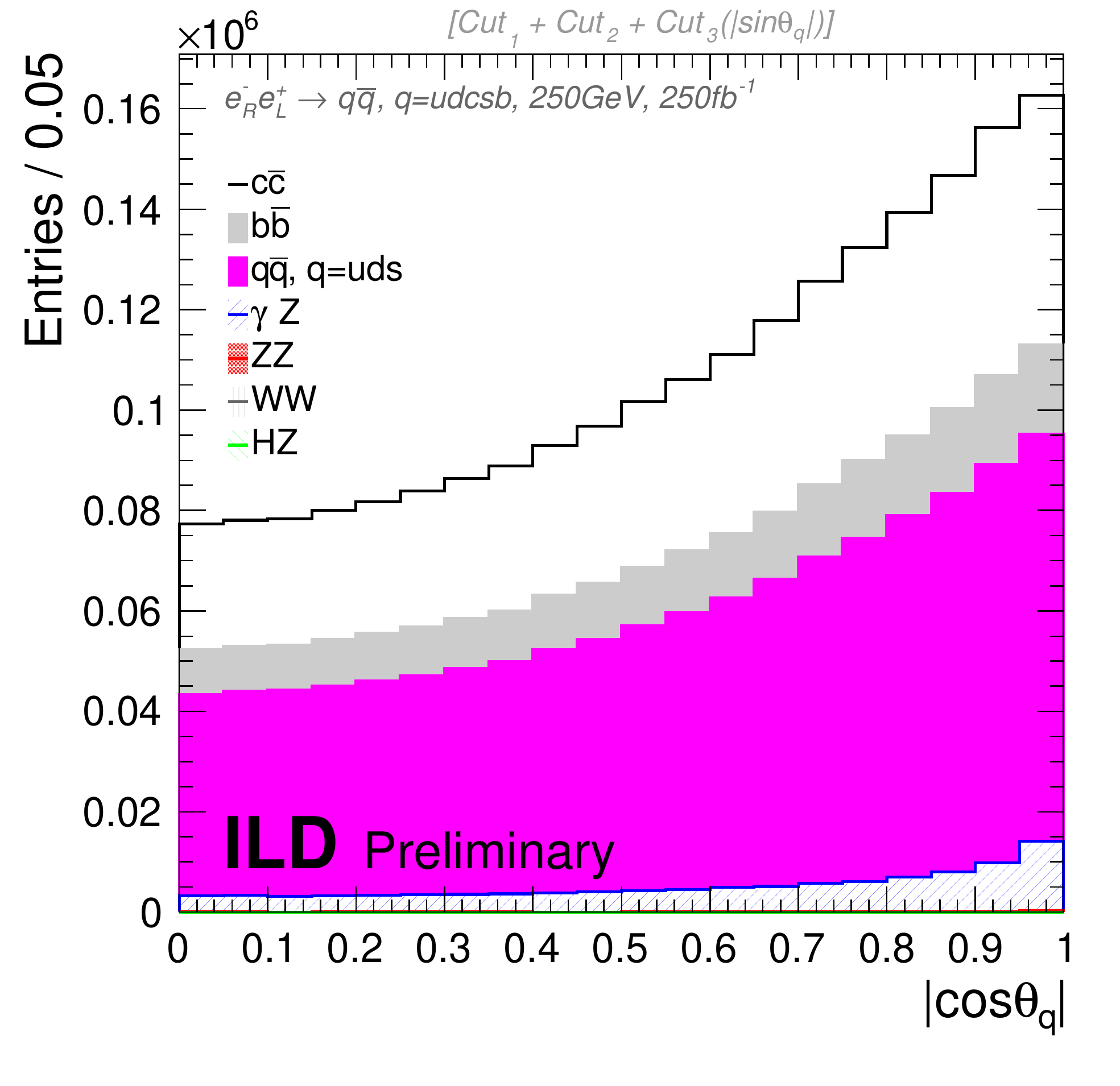} \\
      \includegraphics[width=0.45\textwidth]{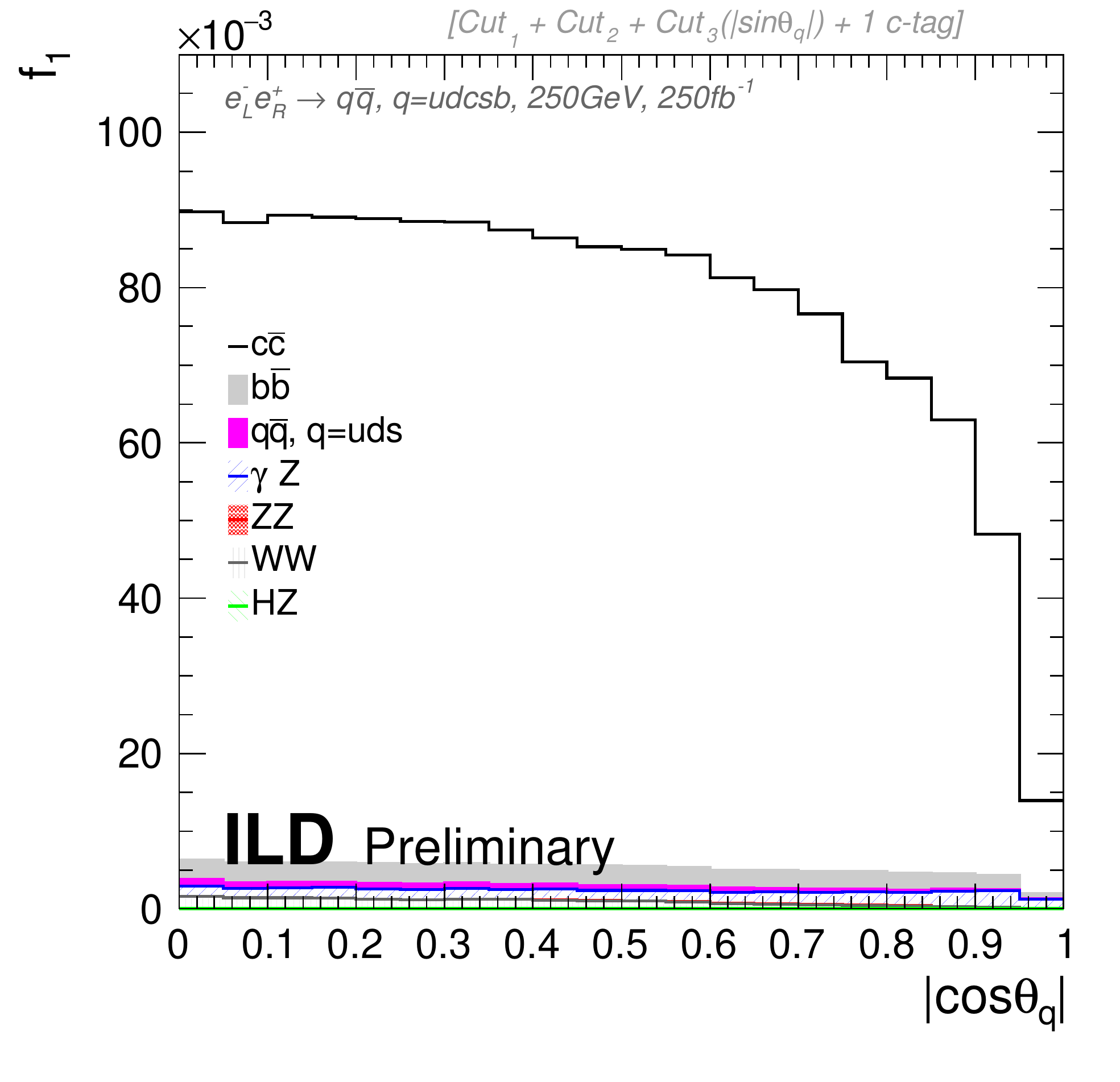} &
      \includegraphics[width=0.45\textwidth]{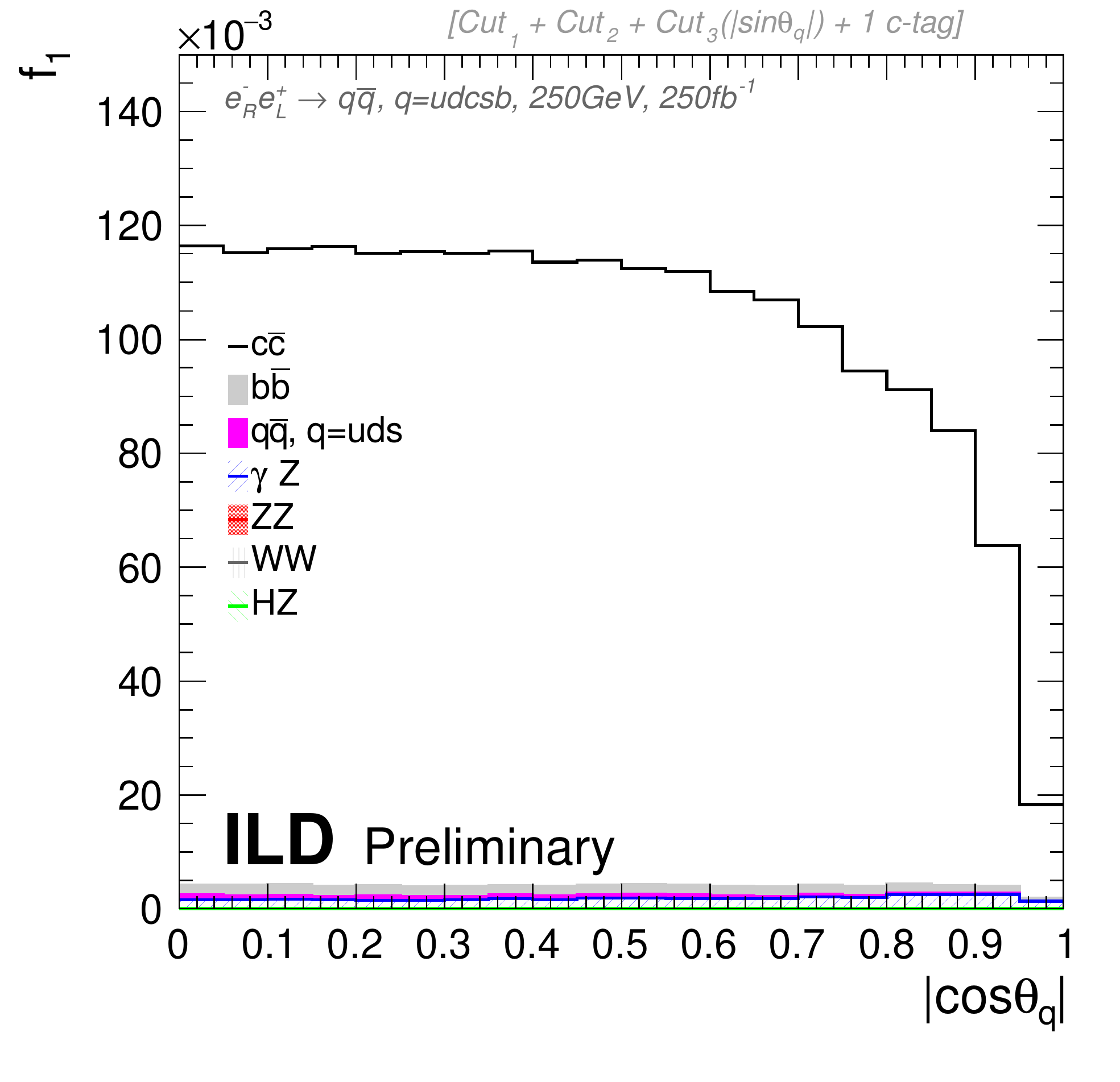} \\
      \includegraphics[width=0.45\textwidth]{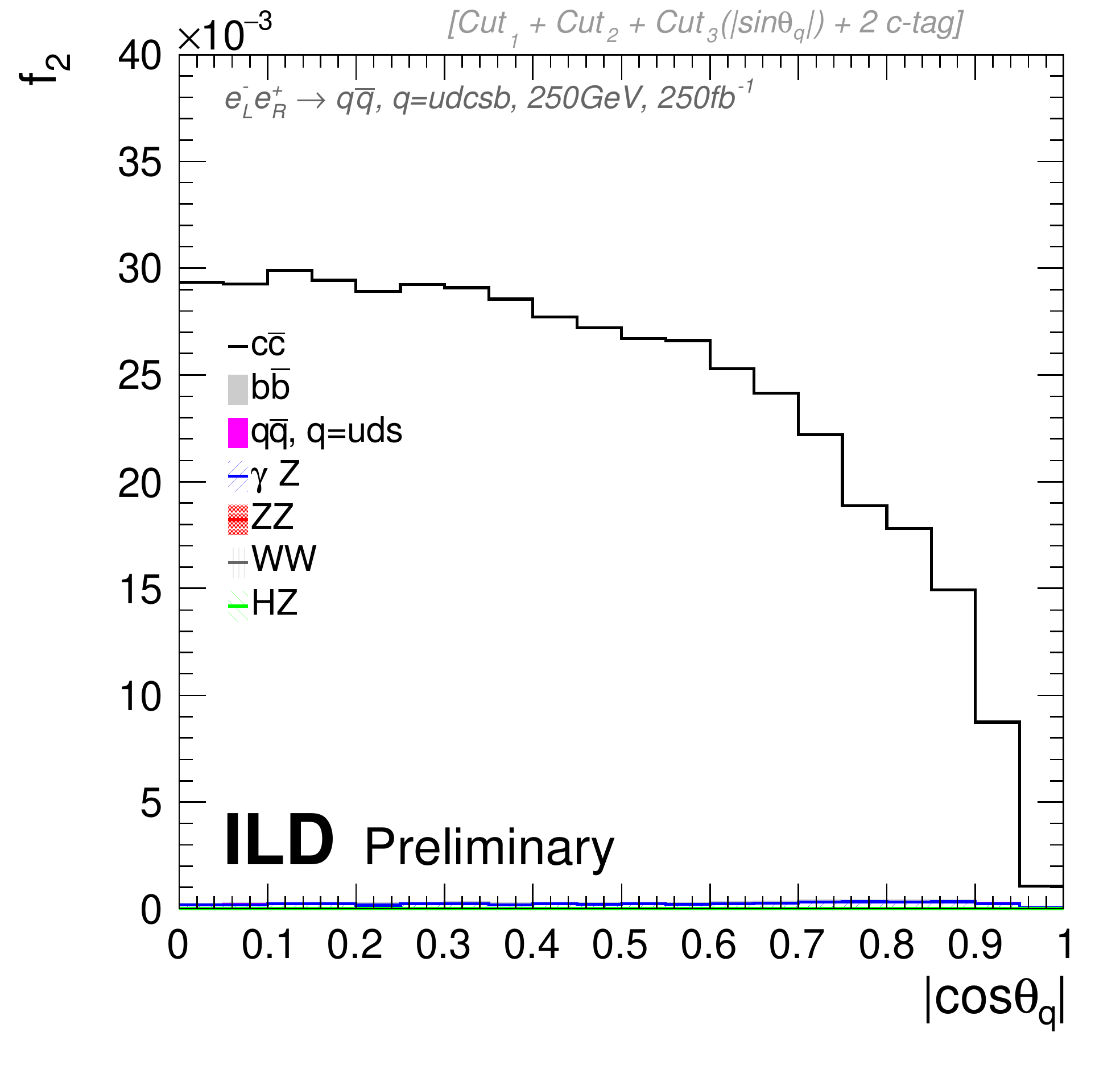} &
      \includegraphics[width=0.45\textwidth]{f2_eL-eps-converted-to.pdf} \\
    \end{tabular}
\caption{First row: distribution of preselected 2-jet distributions. Second row: the $f_{1}$ distribution as defined in Eq. \ref{eq:Rc}. Third row: the $f_{2}$ distribution as defined in Eq. \ref{eq:Rc}. All distributions are shown for the 100\% left and right handed polarisarion cases. \label{fig:selection2}}
\end{center}
\end{figure}

The reconstructed $f_{1}$ and $f_{2}$ distributions are shown in the second and third rows of Figure \ref{fig:selection2}.
The angular distribution of the estimated $\epsilon_{c}$ and $\rho_{c}$ can be seen in Figure \ref{fig:rho} (red graphs).

\subsection{Experimental systematic uncertainties}

The tagging efficiency $\epsilon_c$ is the main source of uncertainty on the determination of \Rc,
and we find that it can be determined at the per mile level. The other systematic uncertainties
considered are close to negligible.

\subsubsection*{Correlation factor}

The correlation factor is compatible with zero for most values of \costheta and therefore it
will not affect to the measurement of \Rc and $\epsilon_{c}$.
It is important to remark that
for the measurements made at LEP1 and SLD, this factor was of the order of the few percent
and had a significant impact on the
systematic uncertainty not only for the $c$-quark case, but also for the $b$-quark case.
Due to the small size of the beam spot expected at the ILC
and the proximity to the beam line of the first tracking layers in the ILD,
the correlation effects will be very much reduced. 

For 250 GeV collisions, \Rb is more or less a factor two lower than \Rc and it 
will be measured at the per mile level \cite{epsproc}, when recorded 2000 fb$^{-1}$ of luminosity.

\subsubsection*{Tagging and mistagging efficiencies}

The high purity of the $c$-tagging of ILD corresponds to small values of $\epsilon_{b}$ of 1.3\%. 
Using control samples (for example, considering only events in which one jet is tagged as a $b$-quark jet with high purity)
this value could be measured to better than 1\%, assuming a
factor two improvement from \cite{Abe:2005nqa}.
For the lighter quarks, $\epsilon_{uds}$, is almost negligible $\epsilon_{uds}=0.1\%$
and has a minimal impact on the extraction of $\epsilon_{c}$ and \Rc.
It can be extracted from MC: current knowledge of the $g\rightarrow c\bar{c}$ coupling is
known to 12\%\cite{ALEPH:2005ab}.
The value of both mistagging efficiencies $\epsilon_{b}$ and $\epsilon_{uds}$ should also account
for the small differences observed in the preselection between flavours, due to the quark mass effects.
These differences are of the order of the 1-2\% and are estimated to be known to 1-10\% \cite{Abreu:1997ey,rodrigomb}.
Therefore this effect will affect minimally the determination of the different $\epsilon_{c}$.

\subsubsection*{Background subtraction}

In contrast with previous experiments where the collisions
occurred at the $Z$-mass resonance dominating the signal, we will have other background contributions
to the preselection that will be of the order of the 20\% of the total
\ccbar signal. The dominating one will be the radiative return. We expect to understand this background
to better than 1\% level from \eeZqq measurements during the GigaZ physics program
of the ILC \cite{Yokoya:2019rhx,Irles:2019xny}. Therefore, this effect will also have minimal impact on the determination of the different $\epsilon_{q}$.

\subsubsection*{Beam polarisation}

For the estimation of the uncertainties due to the measurement of the beam polarisation, we take
the numbers from \cite{Karl:424633}.

\begin{table}[!ht]
 \begin{center}\renewcommand{\arraystretch}{1.8}
  \scriptsize
  \begin{tabular}{cccc}
    \hline
    \multicolumn{4}{c}{{\bf Beam polarisation uncertainty}}\\
    \hline
    $\Delta P^{-}_{e^{-}}~ [\%]$ & $\Delta P^{+}_{e^{-}}~ [\%]$ &  $\Delta P^{-}_{e^{-}}~ [\%]$ &  $\Delta P^{+}_{e^{+}}~ [\%]$ \\
    \hline
    0.1 & 0.04 & 0.1 & 0.14 \\
    \hline
    \end{tabular}
  \end{center}
 \caption{\label{tab:polerror} Uncertainty on the beam polarisation. Numbers extracted from \cite{Karl:424633} }
\end{table}

Since the total cross section depends on the
polarisation of the beams, it will affect the final result. However, due to the reduced
beam polarisation uncertainties and that the observable is a ratio, the impact is negligible.

\subsection{Results}

After all these considerations, we estimate that the $\epsilon_{c}$ will be determined with a precision of at least
0.2\% for a recorded luminosity of 2000 fb$^{-1}$.
This is translated to the following expectations for the 
precision on \Rc for the two beam polarisations for the ILC250 physics programme:

\begin{equation}
  \begin{aligned}
    \Delta \Rc(\eLpR)= 0.10\% (stat.) + 0.16\% (syst.) \\
    \Delta \Rc(\eRpL)= 0.13\% (stat.) + 0.17\% (syst.) 
    \label{eq:Rcresults}
  \end{aligned}
\end{equation}

for a total luminosity of 2000 fb$^{-1}$.

\begin{figure}[!h]
\begin{center}
    \begin{tabular}{cc}
      \includegraphics[width=0.45\textwidth]{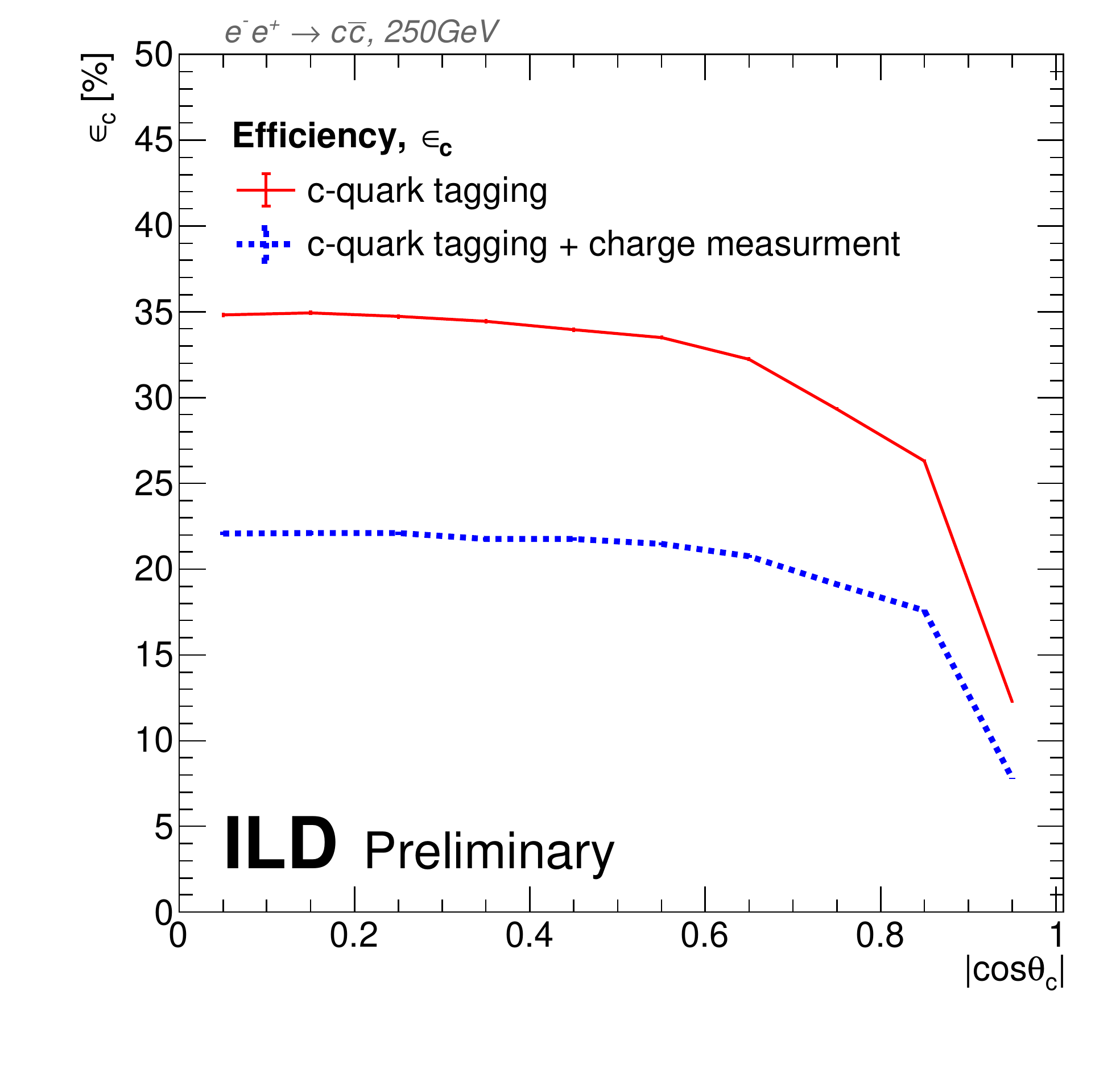} &
      \includegraphics[width=0.45\textwidth]{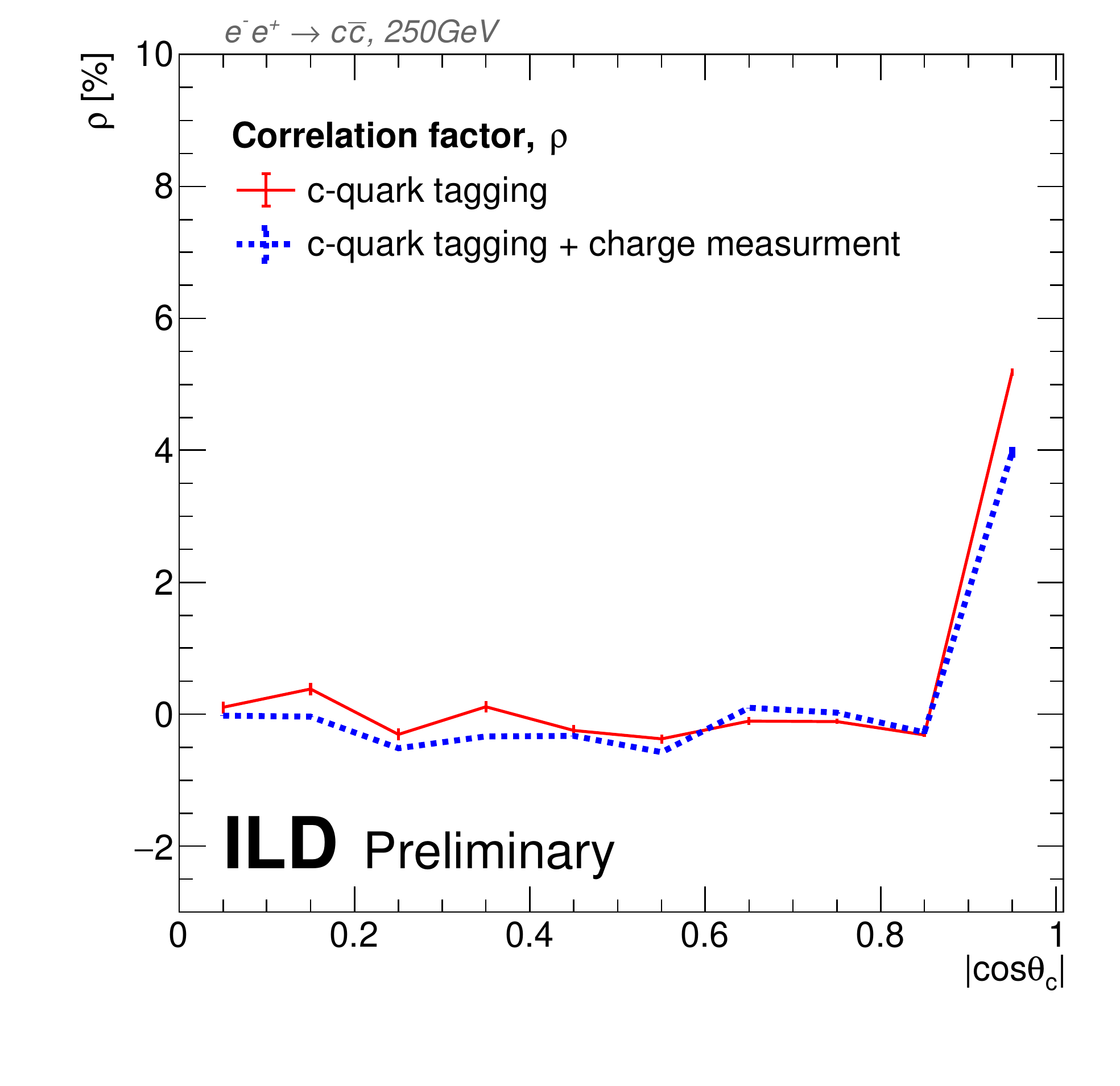} \\
    \end{tabular}
\caption{\label{fig:rho} Efficiency (left) and correlation factor (right) for $c$-quark tagging and $c$-quark charge measurement.}
\end{center}
\end{figure}

\section{Measurement of \Afbc}

After the selection of a highly pure \ccbar sample,
we need to determine the  jet charges to
reconstruct the angular distribution.
The $c$-quarks mostly produce $D^{0}/D^{\pm}/D_{s}$-mesons.
The decay branching ratio of $D^0$ to charged kaons is $\sim50\%$.
For the $D_{s}$ this number is somewhat lower: $\sim33\%$.
The $D^{\pm}$ produce one and three prongs in their decays,
with only $\sim30\%$ of the cases having a charged kaon in the final state.
In all cases, identifying a kaon in a secondary vertex gives
direct information on the charge of the original $c$-quark.
This method of quark charge determination is called {\it Kaon-}method.
If the Kaon is not identified, the total charge of the reconstructed secondary vertex is used.
This is specially interesting for the decays involving $D^{\pm}$-mesons.
This second method is called {\it Vtx-}method.
To apply these methods, we require excellent tracking, vertexing
and particle identification capabilities.
While ILD has a 99\% probability to reconstruct the relevant charged tracks, this probability falls to 96\% for
tracks connected to the micro-vertex and having a significant offset. Revisiting the tracking and vertexing algorithms to 
improve this value is an ongoing activity in the collaboration. This probability drops rapidly for jets reconstructed
at $|\costheta|>0.9$ due to imprecise track reconstruction outside of the acceptance of the micro-vertex detector. 
Modifying this geometry
and/or prolonging the first layer of the barrel is under consideration by the collaboration.
Due to the high granularity of the ILD TPC, the power of separation of kaons and other hadrons is large
enough to provide kaon identification with high efficiency and purity for jets with $|\costheta|<0.9$ and 
momentum above 3 GeV. 
For this process, kaons are identified with a purity reaching 90\% at 88\% efficiency.

\begin{figure}[!h]
\begin{center}
    \begin{tabular}{cc}
      \includegraphics[width=0.45\textwidth]{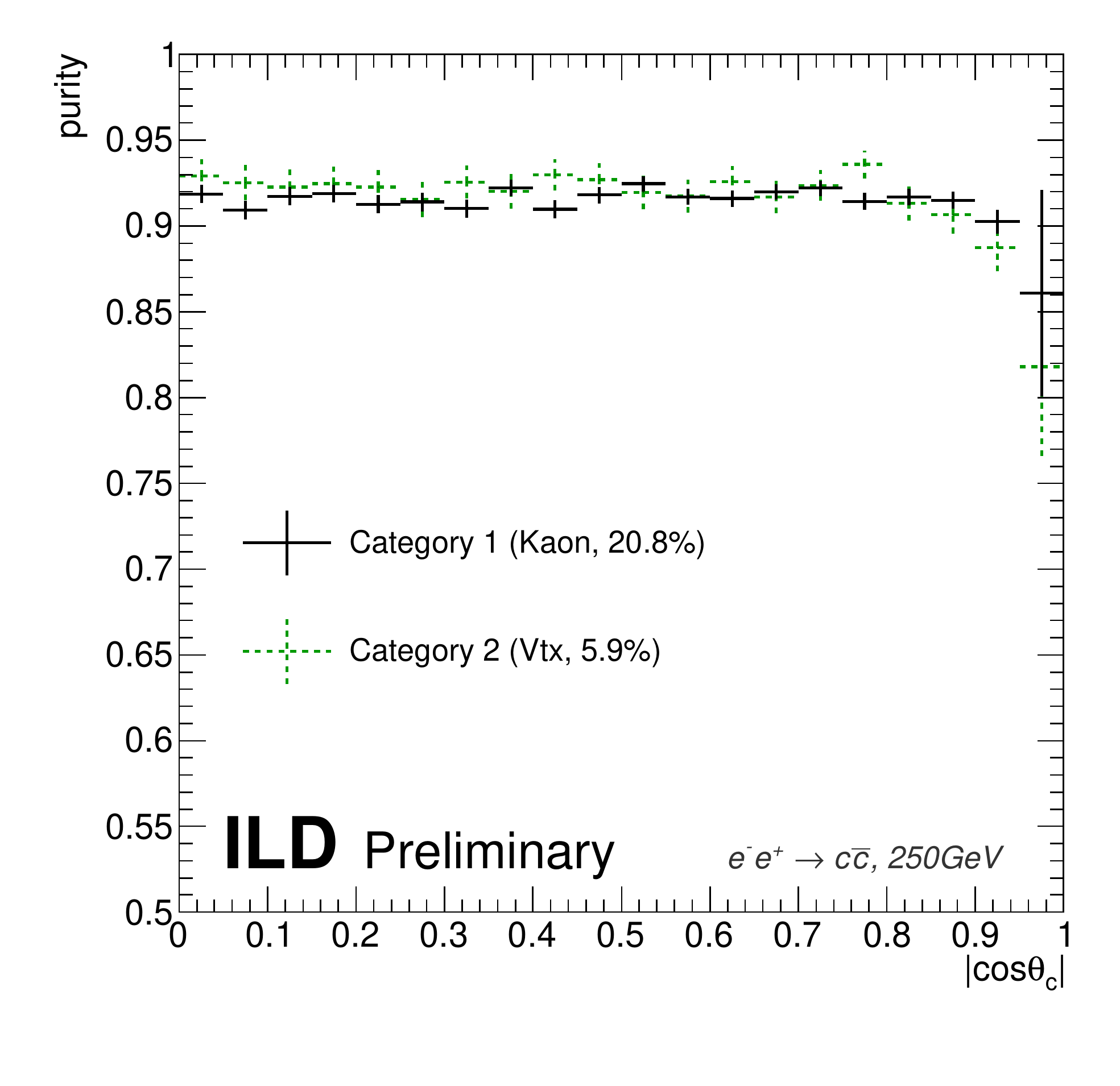} &
      \includegraphics[width=0.45\textwidth]{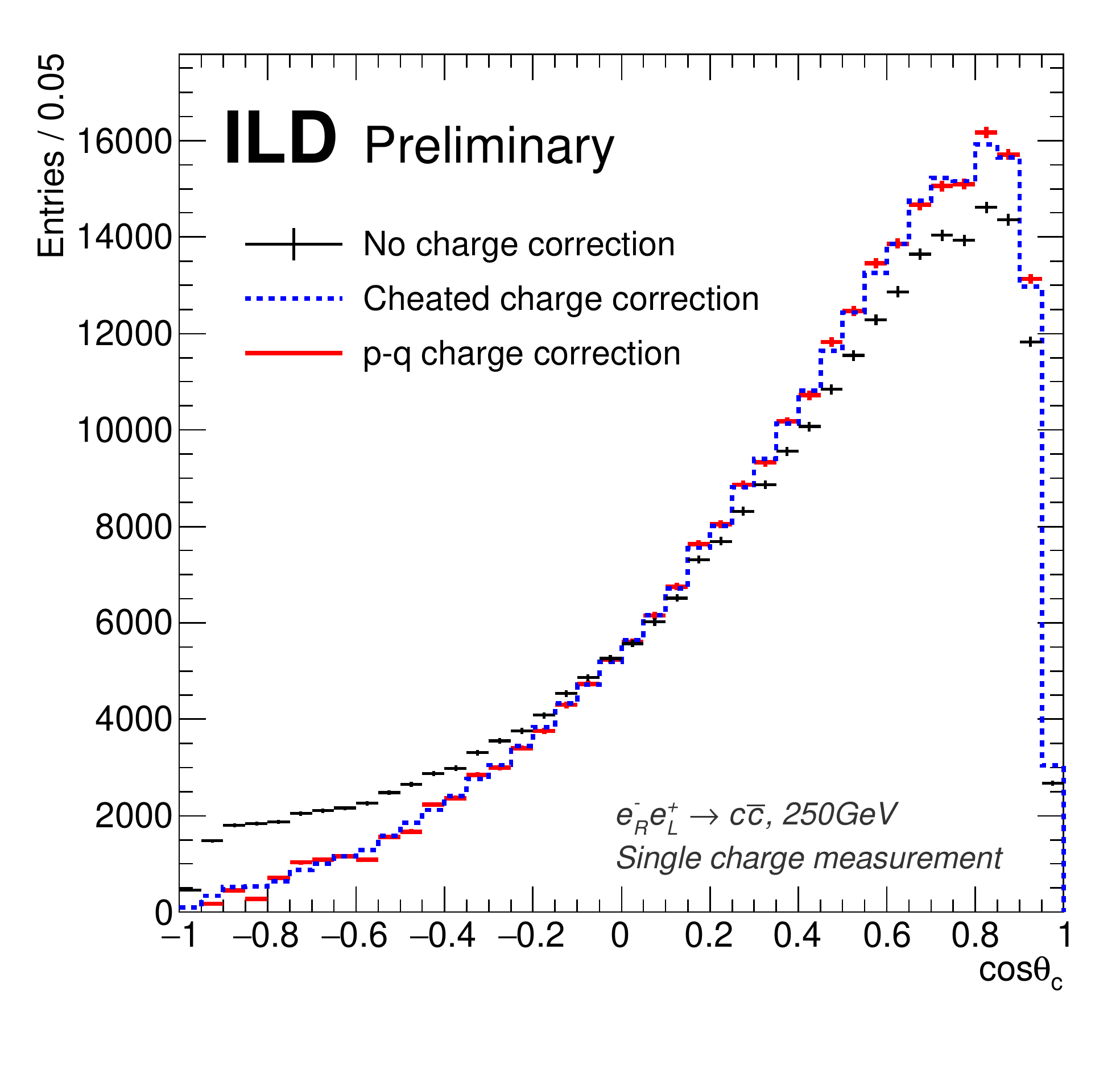} 
    \end{tabular}
\caption{\label{fig:chargecorr} Left plot: Measured purity of each method. Right plot: the performance of the p-q in single events: p-q is calculated using double tagged events but it is applied to single charge events.}
\end{center}
\end{figure}

The charge measurement allows us to attribute a sign to the reconstructed \costhetac value. However a sign-flip may be induced
due the missed tracks in a reconstructed vertex 
and the misidentification of kaons. In order to correct for that, we need to know with high precision
the probability of getting the charge correctly using the different methods. This probability is called purity and
it is measured using the data itself by comparing
two different reconstructed samples in which the charge of both jets has
been measured: the sample in which both jets were estimated to have different/the same charge.
The method is described in detail in \cite{epsproc}. The purity of each method is shown in the left plot of Figure \ref{fig:chargecorr}.
Once that the purity of each method is well known, we can apply it to correct the charge mismeasurements on the reconstructed
distributions. The performance of the method is shown in the left plot of Figure \ref{fig:chargecorr}.

\begin{figure}[!h]
\begin{center}
    \begin{tabular}{cc}
      \includegraphics[width=0.45\textwidth]{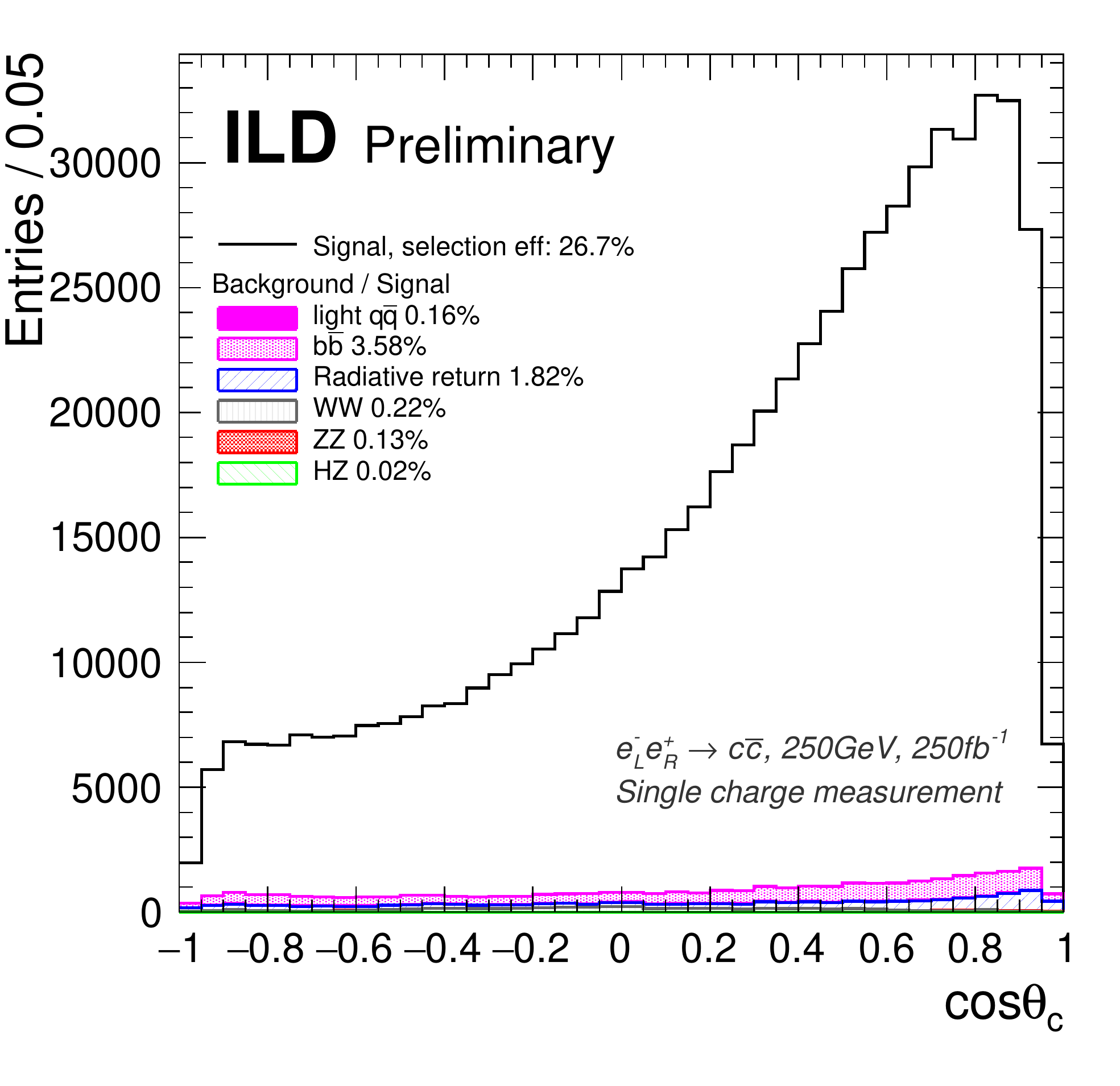} &
      \includegraphics[width=0.45\textwidth]{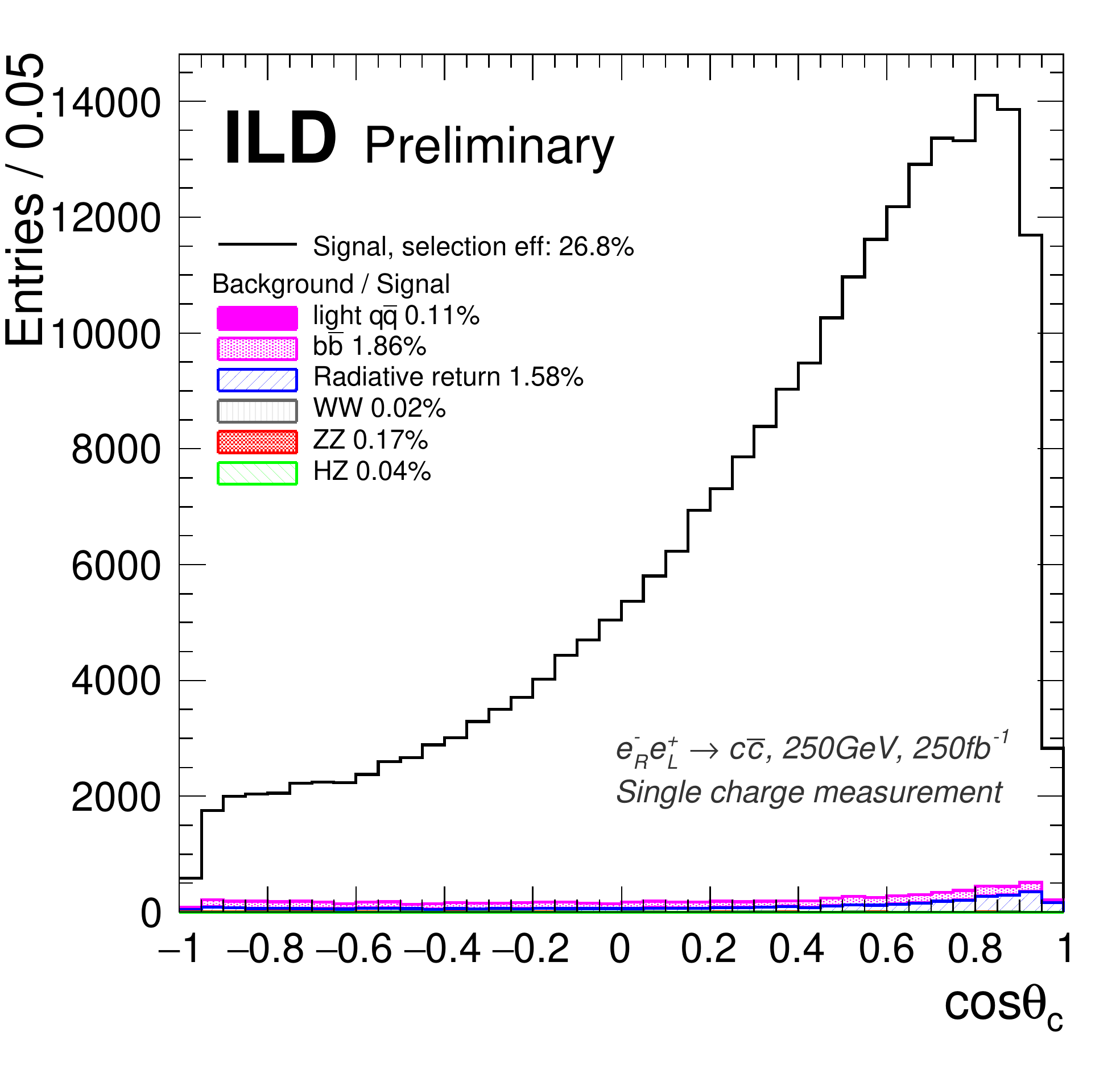} \\
      \includegraphics[width=0.45\textwidth]{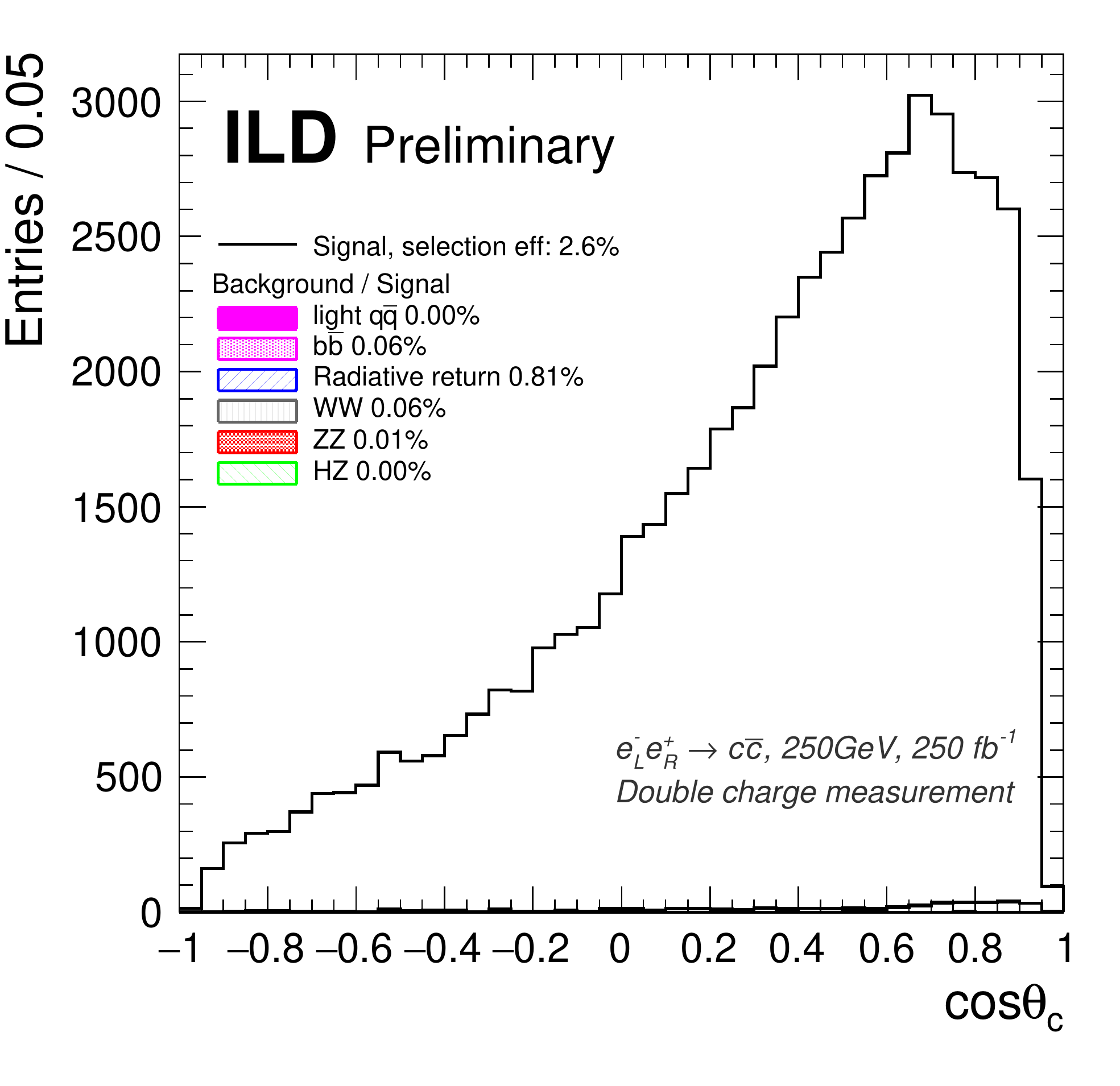} &
      \includegraphics[width=0.45\textwidth]{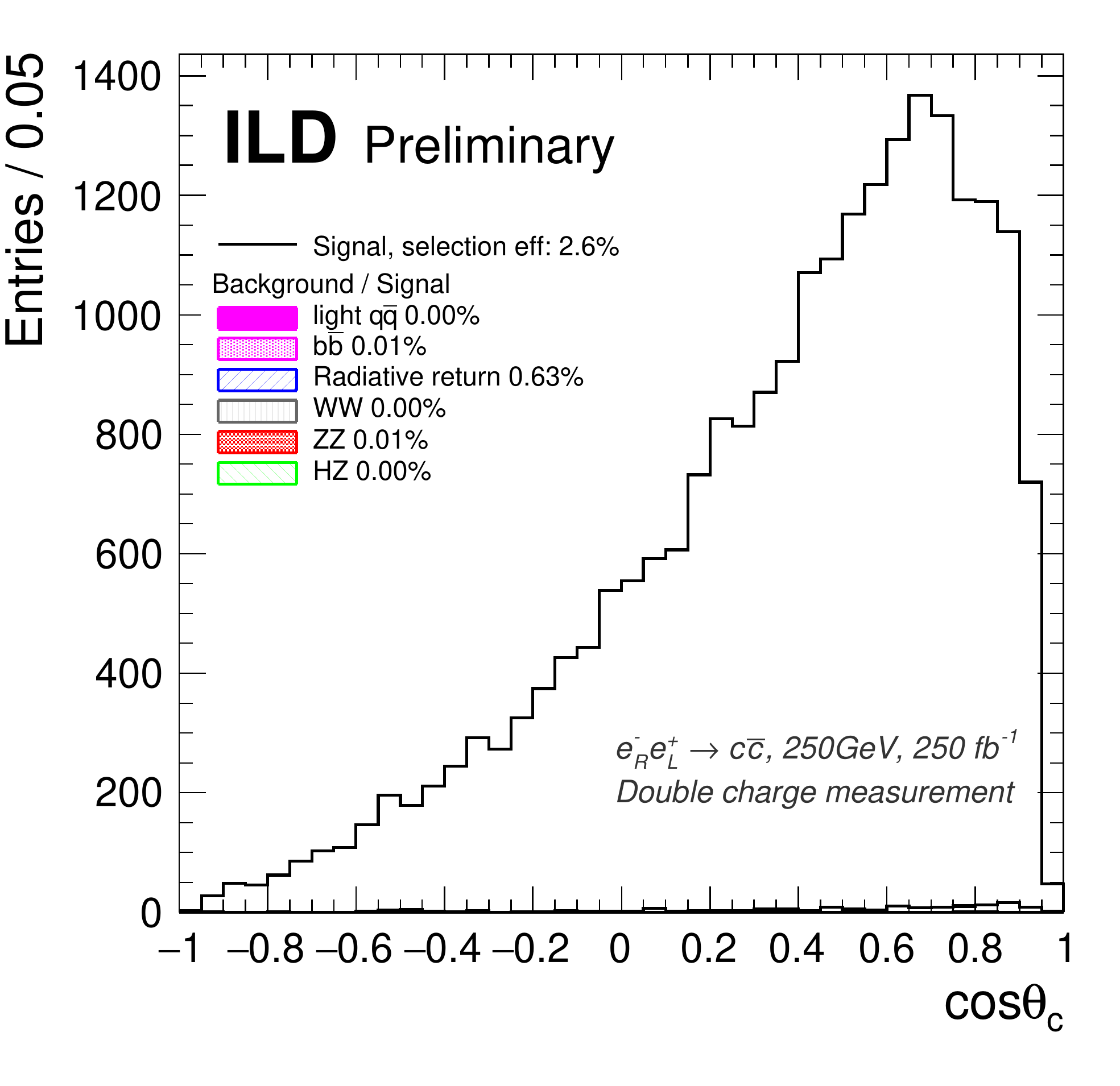} 
    \end{tabular}
\caption{\label{fig:reco} Differential signal and background distributions after the charge calculation using the {\it Vtx-} and {\it Kaon-}method defined above. We compare distributions in which both jets have compatible charges (lower row) or in which only one jet with charge is required (upper row).}
\end{center}
\end{figure}

The final reconstruction efficiency requiring at least one jet with the charge measured is of the
25.7\% (20.8\% using the {\it Kaon-}method and using the 5.9\% {\it Vtx-}method).
The determination of the \Afbc value is done by integrating the measured distribution.
This distribution is shown in Figure \ref{fig:reco} (upper row) for both polarisations.
For completeness, we also show the distributions with double charge measurement.
This is done after the correction of the charge migrations and after the selection efficiency corrections are applied.
The latter is done by measuring the $c$-tagging plus charge calculation efficiency, $\epsilon_{c,charge}$, in the same way
as in the \Rc case, but applying $c$-tagging and charge measurement at the same time.

\subsection{Experimental systematic uncertainties}

The measurement of \Afbc mostly suffers from the same systematic
uncertainties as \Rc. Therefore we will only describe the points
that are specific for this observable.

The expected $\epsilon_{c,charge}$ and the corresponding correlation factor
are shown in Figure \ref{fig:rho} as blue dotted lines
and will be determined with a precision of the order of 0.3\%.
It is important to remark that as the \Afbc is a ratio in which
the tagging efficiency will appear in both the numerator and denominator.
For \Rc, the tagging efficiency (and mistagging efficiencies) appeared only in the
numerator.
Therefore, we are not interested in the global value of $\epsilon_{c,charge}$ but in the relative variations between the correction factors.
In any case, for safety, we will reduce the analysis of the distribution
to values of $|\costheta|<0.75$ before the drop in $\epsilon_{c,charge}$.
There is a second effect that is also corrected for: the inhomogeneities on the preselection efficiency
as a function of $|\costheta|$. This is seen in the left plot of Figure \ref{fig:selection}
where small differences of the $\sim$1\% are observed. Due to the reduced
size of such inhomogeneities, their impact will be minor and will not affect the final result
if they are known at the level of a few percent which will be achievable with controlled samples
or with simulations. 

\subsection{Results}

\begin{figure}[!pt]
\begin{center}
    \begin{tabular}{cc}
      \includegraphics[width=0.45\textwidth]{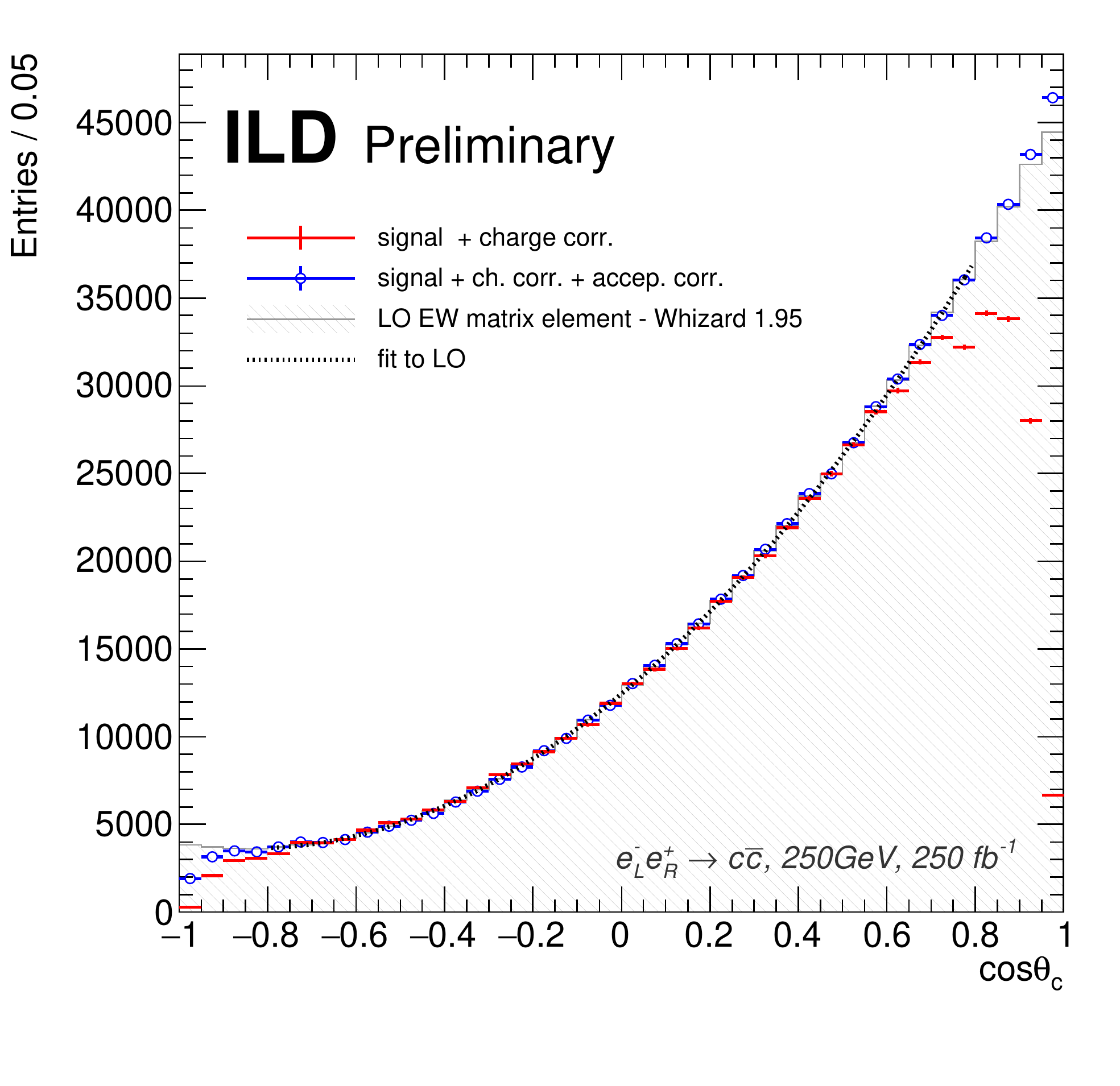} &
      \includegraphics[width=0.45\textwidth]{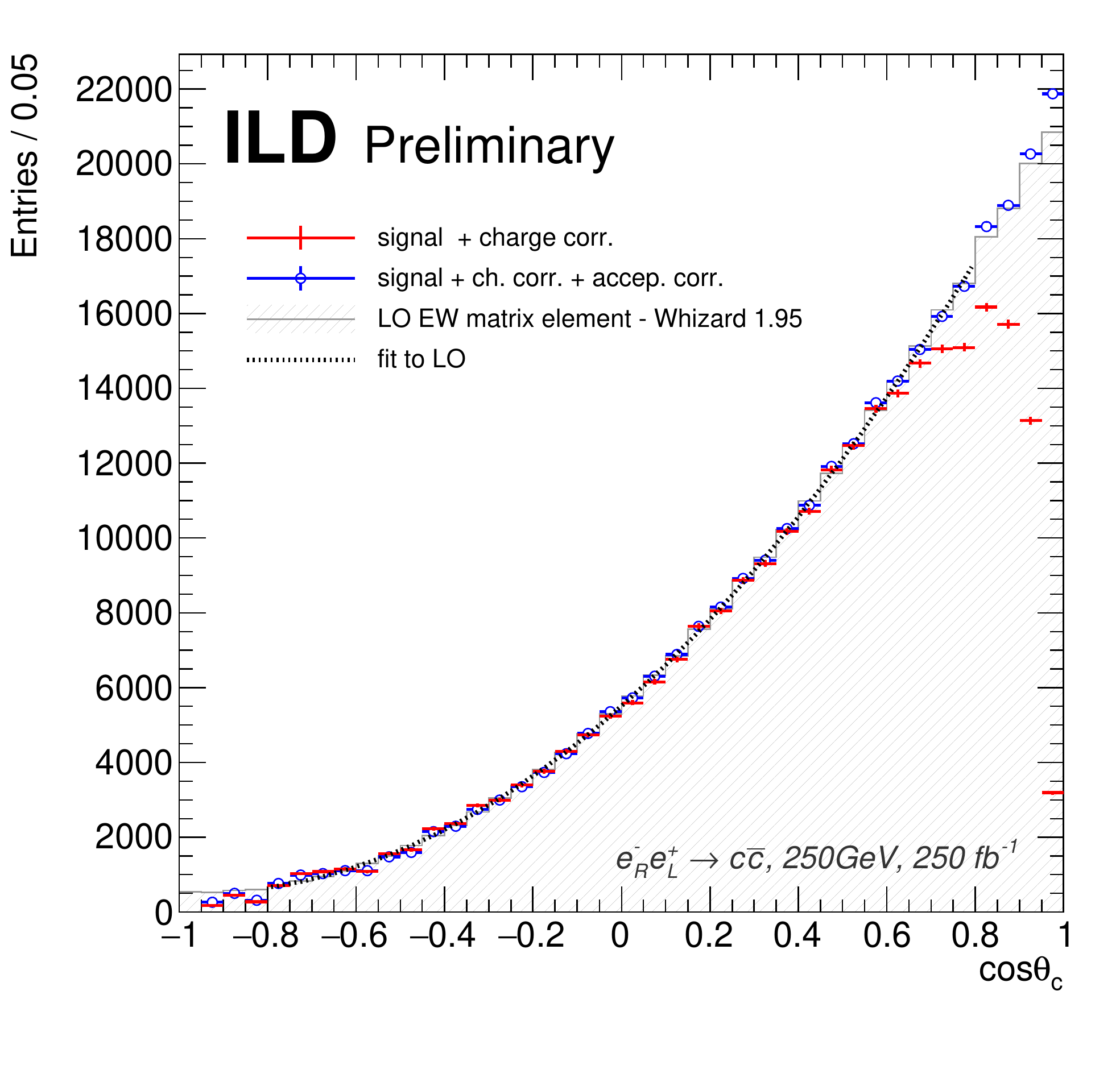} 
    \end{tabular}
\caption{\label{fig:fit} Reconstructed angular distribution for both polarisations. The red empty dots shows the reconstructed distribution before the efficiency and acceptance correction. The blue circles shows the corrected distribution. The shaded grey are shows the prediction at leading order (LO) and the grey curve shows the result of the fit of the LO to the final distribution in a constrained range of \costhetac.}
\end{center}
\end{figure}

The final distributions for both polarisations fitted to the leading-order estimation are shown in Figure \ref{fig:fit}.

The expected precision on the forward-backward asymmetry measurements are, for a recorded luminosity of 2000 fb$^{-1}$:

\begin{equation}
  \begin{aligned}
    \Delta \Afbc(\eLpR)= 0.16\% (stat.) + 0.09\% (syst.) \\
    \Delta \Afbc(\eRpL)= 0.20\% (stat.) + 0.10\% (syst.)  
    \label{eq:Afbcresults}
  \end{aligned}
\end{equation}

\section{Prospects for BSM discoveries}

The results on the expected experimental precisions foreseen for ILC running at 250 GeV are summarised in
the previous sections. For both observables, and both polarisations,
total experimental uncertainties of $\sim0.2\%$ are expected
for the full 2000 fb$^{-1}$ program.
Such accuracies consist a challenge to theoretical high order corrections, particularly
for what concerns the electroweak corrections. It is out of the scope
of this document to discuss this issue.

\begin{figure}[!pt]
\begin{center}
   \includegraphics[width=0.6\textwidth]{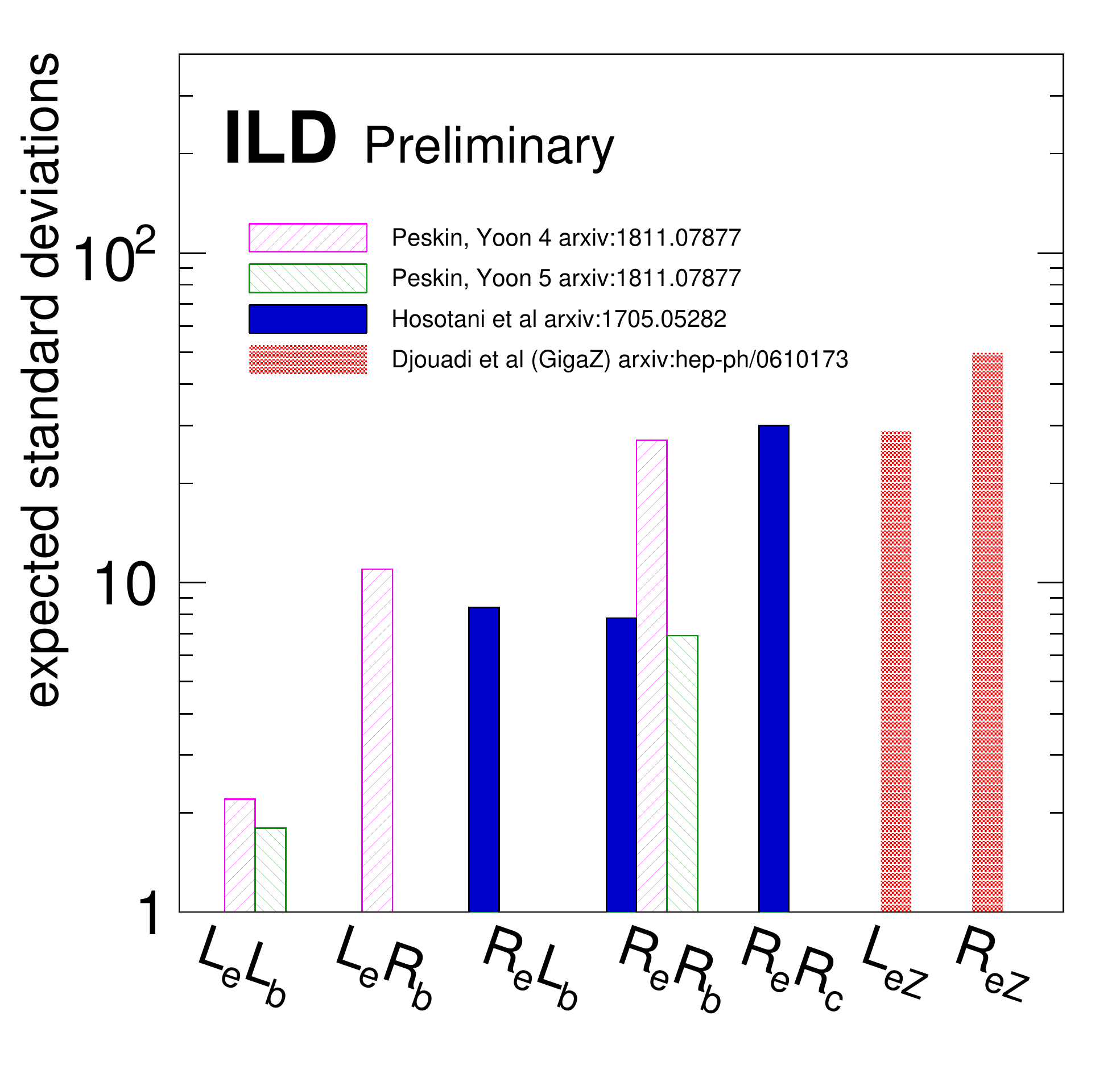} 
   \caption{Expected number of standard deviations for different BSM scenarios when determining the different EW couplings to $c$- and $b$-quark at ILC250.
     The $x$-axis shows the different couplings for different chiralities, following the notation from \cite{epsproc}.
     The expectations for different BSM scenarios are shown: 1) for Djouadi \cite{Djouadi:2006rk} one assumes $m_{Z^{\prime}}=3$ TeV; 2) for the Peskin {\it et al.}
     model \cite{Yoon:2018xud}, two versions are given, labelled as Peskin 4 and Peskin 5; 3) for Hosotani et al.\cite{Funatsu:2017nfm} one assumes	
     $m_{Z^{\prime}}\sim8$ TeV for the 3 resonances.
     These prospects assume the input from
     the ILC GigaZ programme running at the $Z$-Pole \cite{Irles:2019xny} in order to
     improve by a factor $\sim$5 the current precision on the SM $Z$-boson couplings to the different quarks
     measured at the $Z$-pole.	
   } \label{fig:couplings}
\end{center}	
\end{figure}

Many beyond standard models with extended gauge structures 
\cite{Djouadi:2006rk,Funatsu:2017nfm,Yoon:2018xud} predict
large corrections for the standard model electroweak couplings. Therefore,
these models predict large
modifications of the forward backward asymmetry and the cross section.
Some of these models, for example \cite{Funatsu:2017nfm}, predict 
that such kind of effects for all fermions (not only for the heaviest).
The unprecedented precision that
will be achieved at the ILC allows to deeply investigate all these models.
Of particular importance is the fact that - thanks to the beam polarisation at the ILC - we
could inspect the different helicity amplitudes
in order to disentangle between the different models.
The expected precision on the determination of the helicity amplitudes
can be compared with the SM and any BSM predictions. This is done in Figure \ref{fig:couplings}
for several models, extracting the helicity amplitudes from the \Rc, \Afbc and \Rb, \Afbb \cite{epsproc}.
It shows that a modest energy
of 250 GeV, the ILC has a reach which extends well aboce LHC direct searches.
For what concerns the Hosotani model \cite{Funatsu:2017nfm}, the large effect seen for a 8 TeV resonance indicates that
ILC250 can extend its sensitivity even far beyond. In the absence of a signal, one would conclude that
such a Z$^{\prime}$ is heavier than 34 TeV at 95\% C.L.

\section{Summary}

This document summarises the results of a realistic analysis based on full detector simulation and reconstruction
of \eecc processes at the ILC. The results show a large
improvement on the reachable precisions compared with previous experiments. 
The measurement requires determining the charge of both jets identified as originated by a $c$-quark.
This is possible thanks to expected exceptional vertexing capabilities of the ILD
and the charged kaon identification provided by the \dedx information of its high granularity TPC. 
Also highlighted is a major advantage compared with other experiments: the power of separation
and the independent determination of the left and right handed components of the electroweak couplings
thanks to the beam polarisation.

\section*{Acknowledgements}
We would like to thank the LCC generator working group and the ILD software working group for providing the simulation
and reconstruction tools and producing the Monte Carlo samples used in this study.
This work has benefited from computing services provided by the ILC Virtual Organisation, supported by the national resource providers of the EGI Federation and the Open Science GRID.

\section*{References}
\bibliographystyle{JHEP}
\bibliography{bbbar_references} 

\section*{Appendix: Preselection}
\label{app:sel}

\begin{figure}[!h]
  \centering
      \begin{tabular}{cc}
        \includegraphics[width=0.45\textwidth]{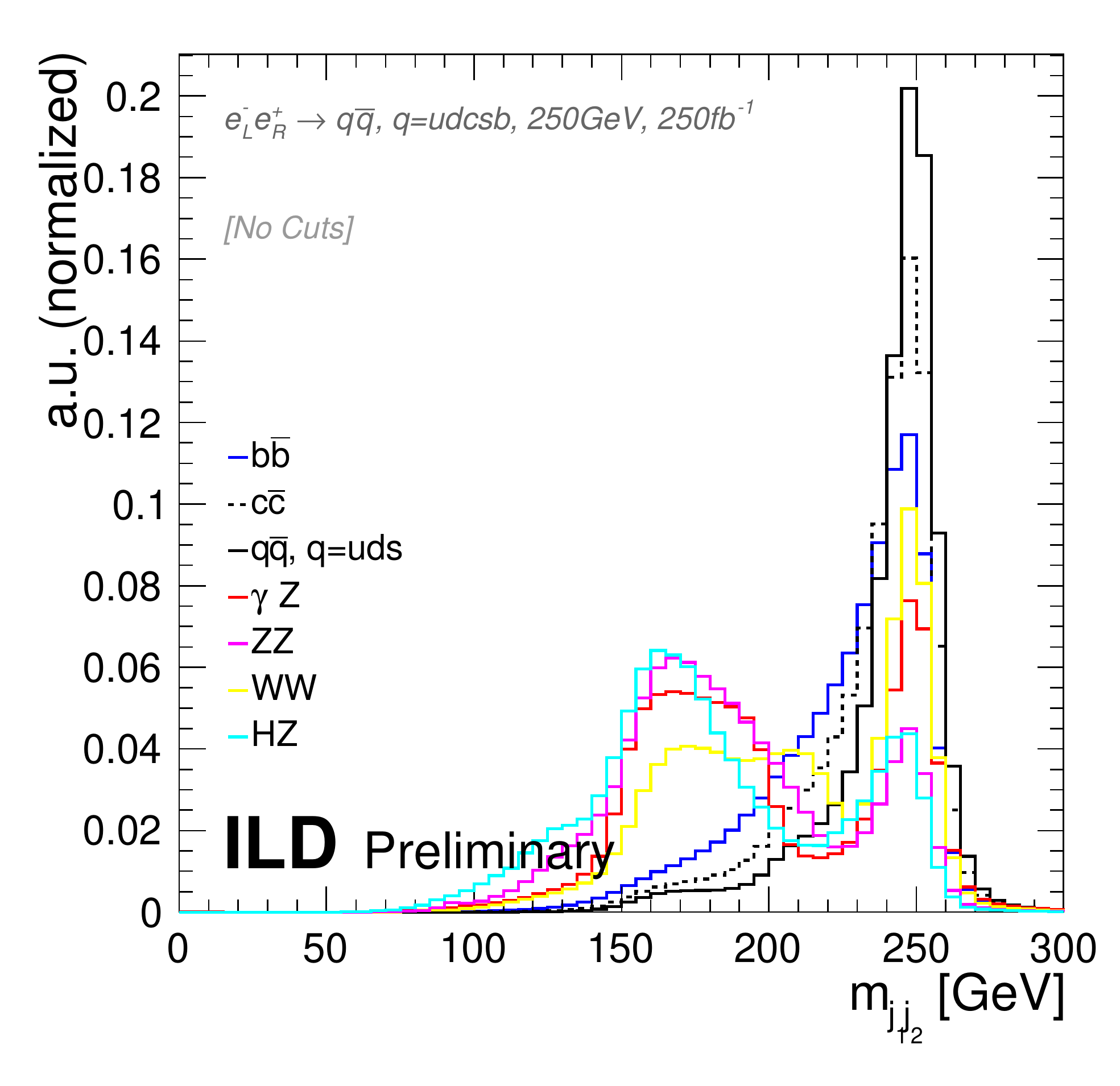} &\includegraphics[width=0.45\textwidth]{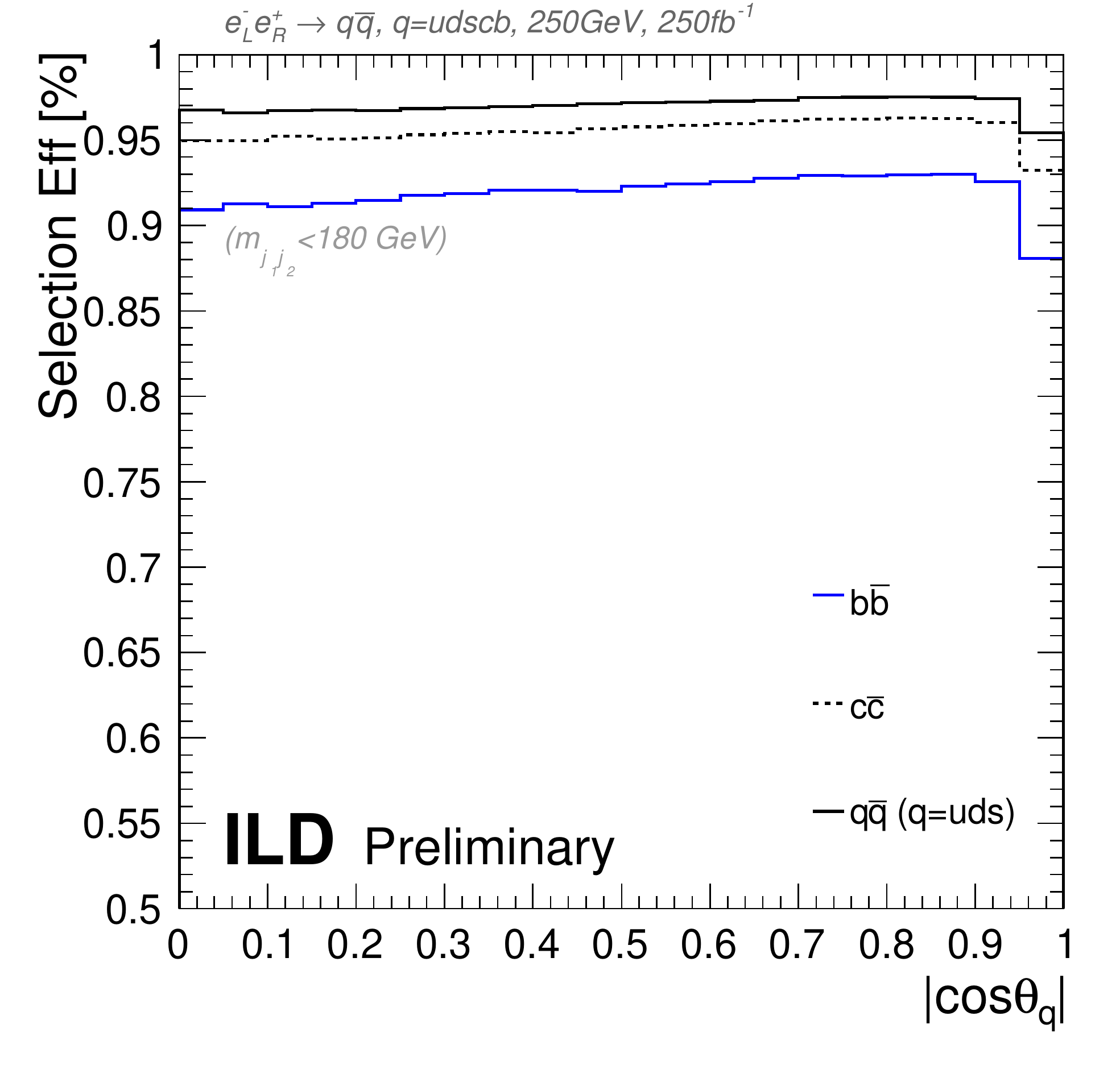}
      \end{tabular}
  \caption{\label{fig:invmass} Invariant mass distribution}
\end{figure}

The event pre-selection proceeds as follows: we reconstruct events with two jets using the Durham algorithm.
We could apply a cut in the invariant mass of the two jet system, in order
to remove the most dominant background: radiative return to the $Z$-pole through ISR.
However, this cut in the invariant mass would introduce a large difference on the preselection
of the different quark flavours since the tails of the distribution highly depends on the quark flavour.
This is mainly associated to the presence of neutrinos in the hadronisation and decay process, which is more
common for heavy than light quarks. The invariant mass distribution of the two reconstructed jets
can be seen in Figure \ref{fig:invmass}. 

\begin{figure}[!h]
  \centering
      \begin{tabular}{ccc}
        \includegraphics[width=0.35\textwidth]{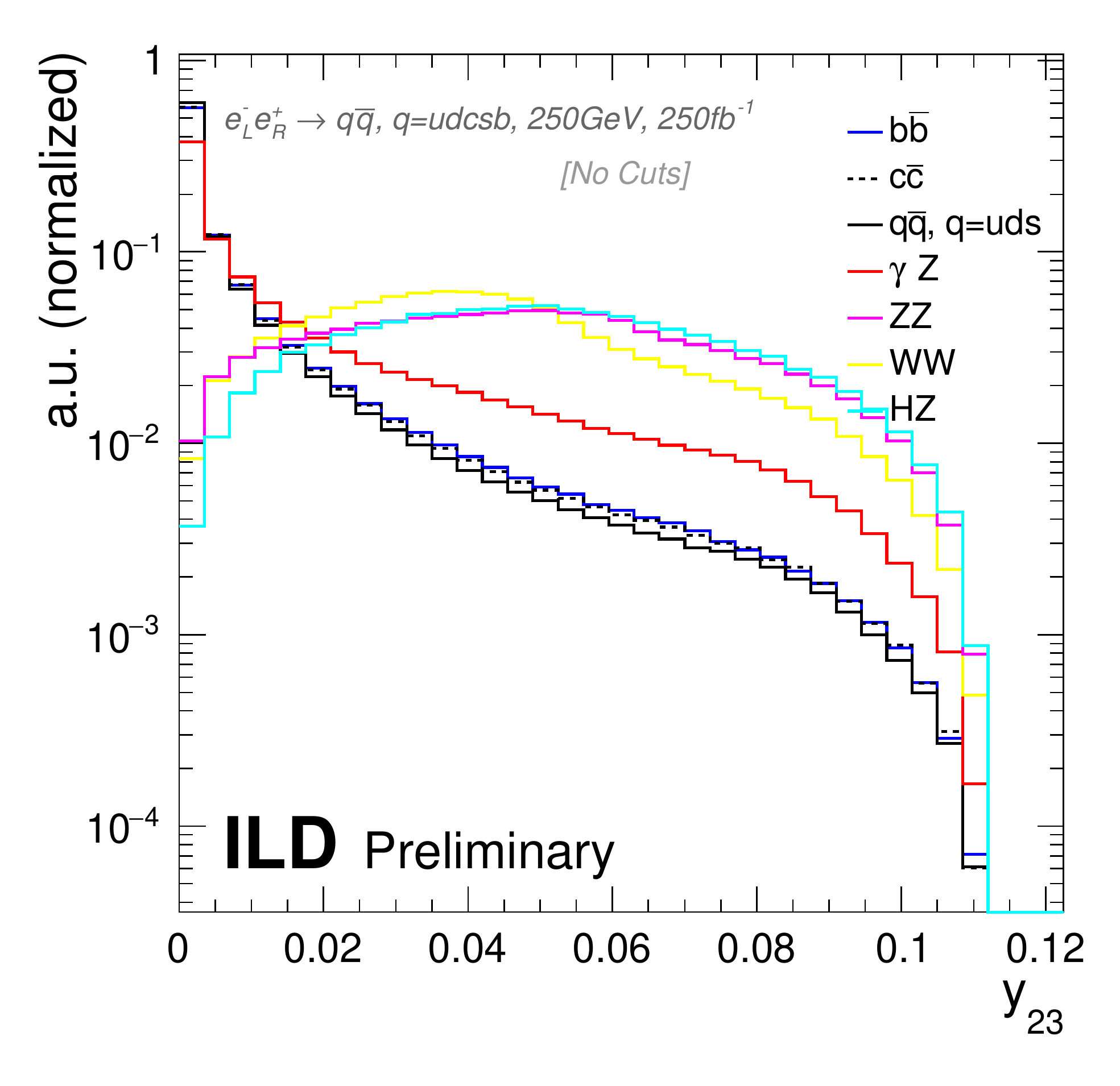} &
        \includegraphics[width=0.35\textwidth]{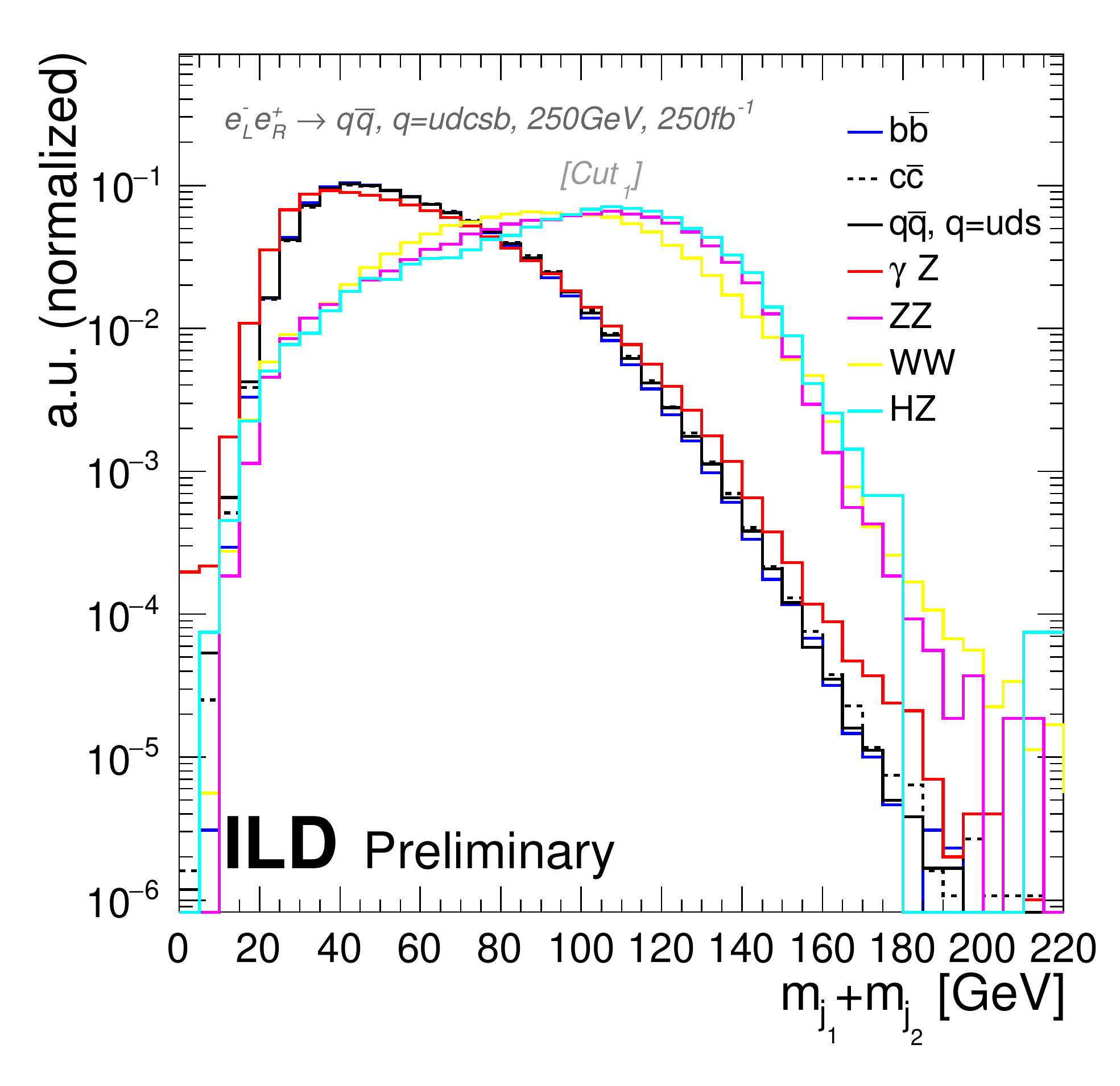} &
        \includegraphics[width=0.35\textwidth]{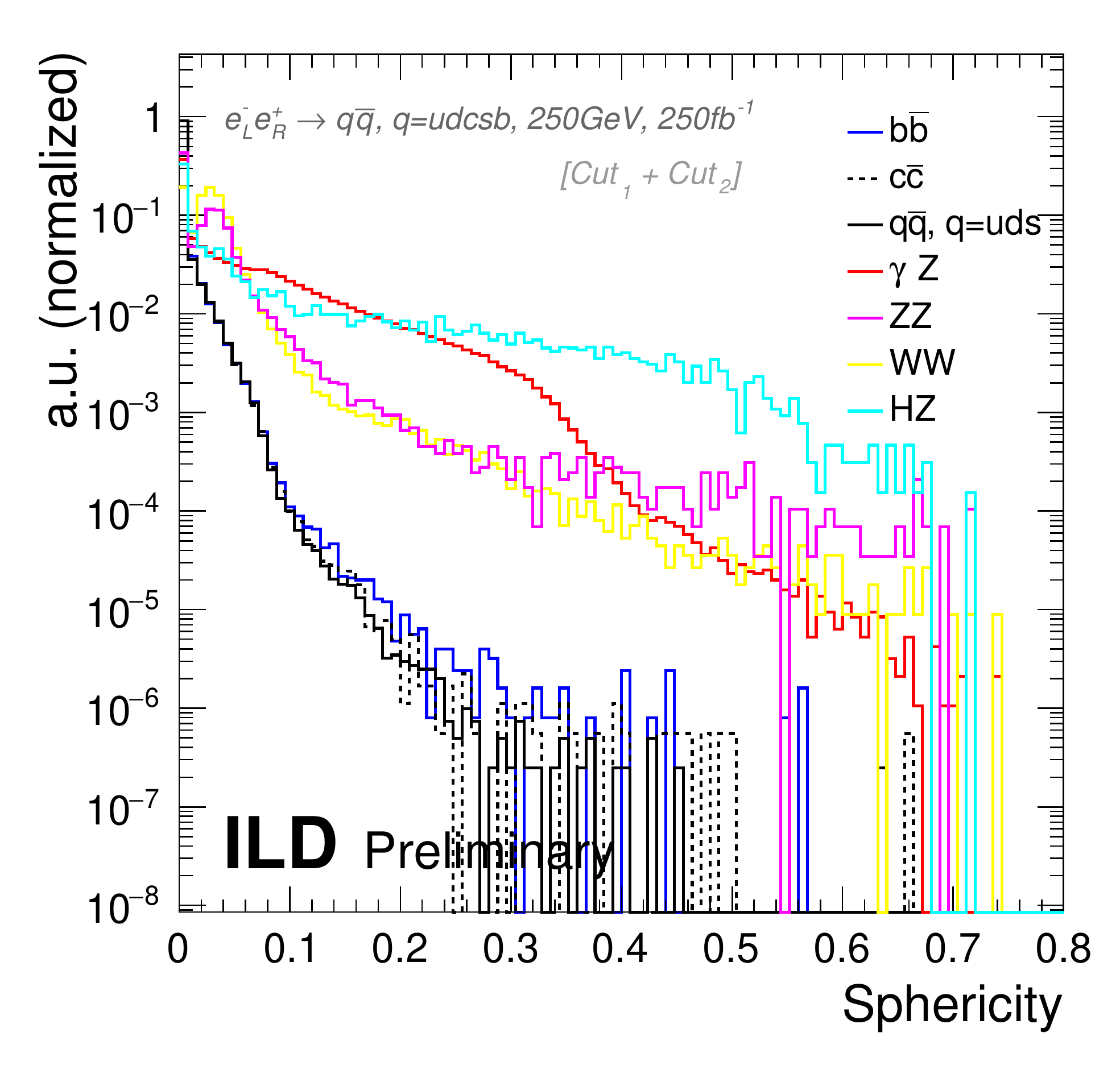} 
      \end{tabular}
  \caption{\label{fig:selection3} Topological variables used for the event preselection.}
\end{figure}

To avoid the issue discussed above, we use topological variables instead of purely kinematic quantities.
The first variable to be used is the Durham-distance $y_{23}$. The distribution
of $y_{23}$ for the different signals and backgrounds is shown in
Figure \ref{fig:selection3}, left plot.
This variable corresponds to the jet clustering distance cut, as defined
by the Durham algorithm, at which a two jet system would be reconstructed as a three jet system. A cut of $y_{2}<0.02$
is applied. This cut introduces a residual flavour dependence due to the differences on QCD FSR due to the
quark masses: the larger is the quark mass, the harder and less collinear is the QCD FSR therefore and therefore
the higher is the possibility
to have a 3-jet like event. 

Further,
a cut in the sum of the two jet masses is applied:
if the sum of the two jet masses is greater than 100 GeV, the event is rejected. This cut helps 
reducing the impact of QCD final state radiation that dilutes
the back-to-back configuration of the two jets
and also helps suppressing the remaining background from ZZ events. See Figure \ref{fig:selection3} middle plot.

\begin{figure}[!h]
  \centering
    \begin{tabular}{cc}
        \includegraphics[width=0.45\textwidth]{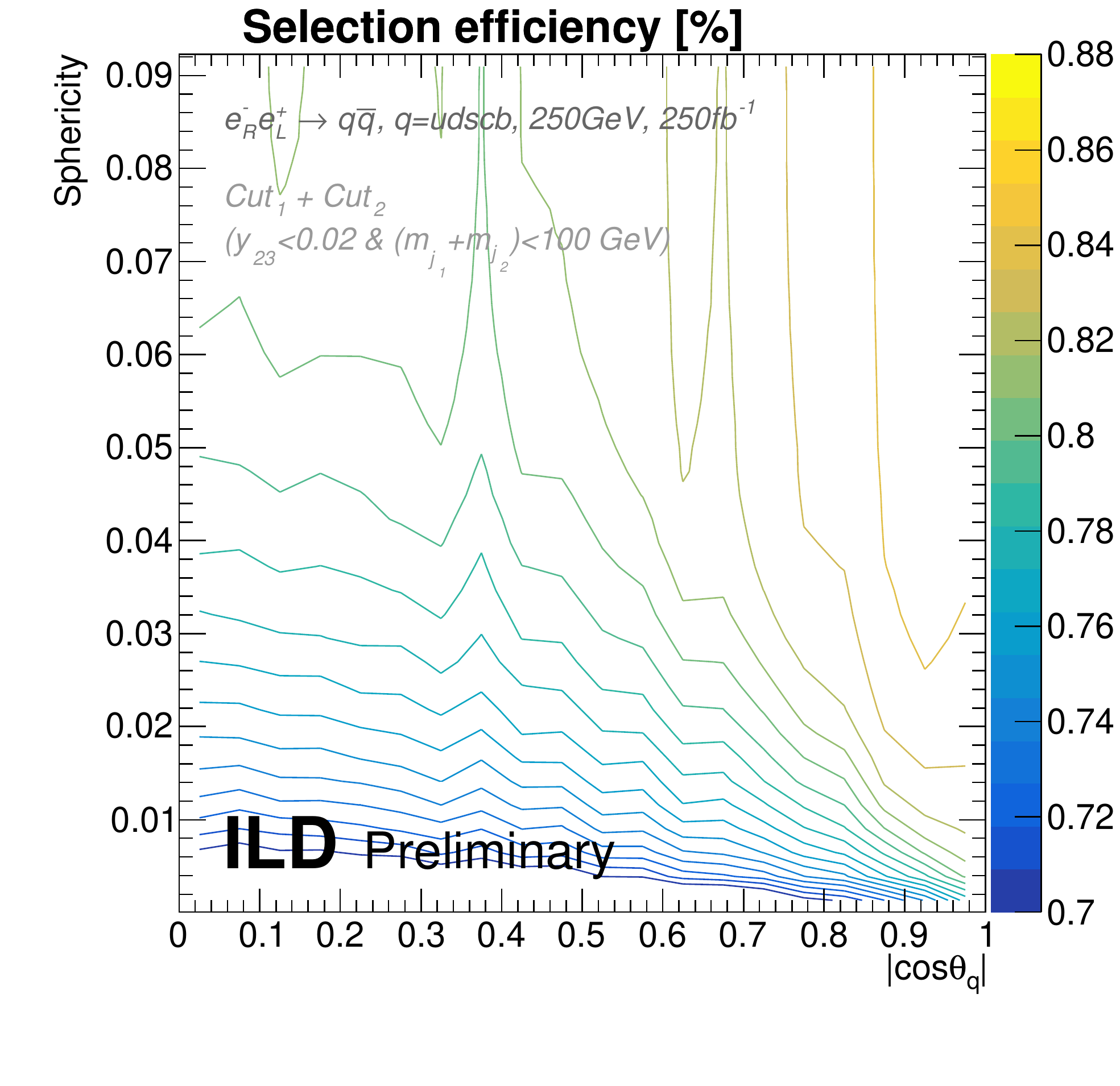} & \includegraphics[width=0.45\textwidth]{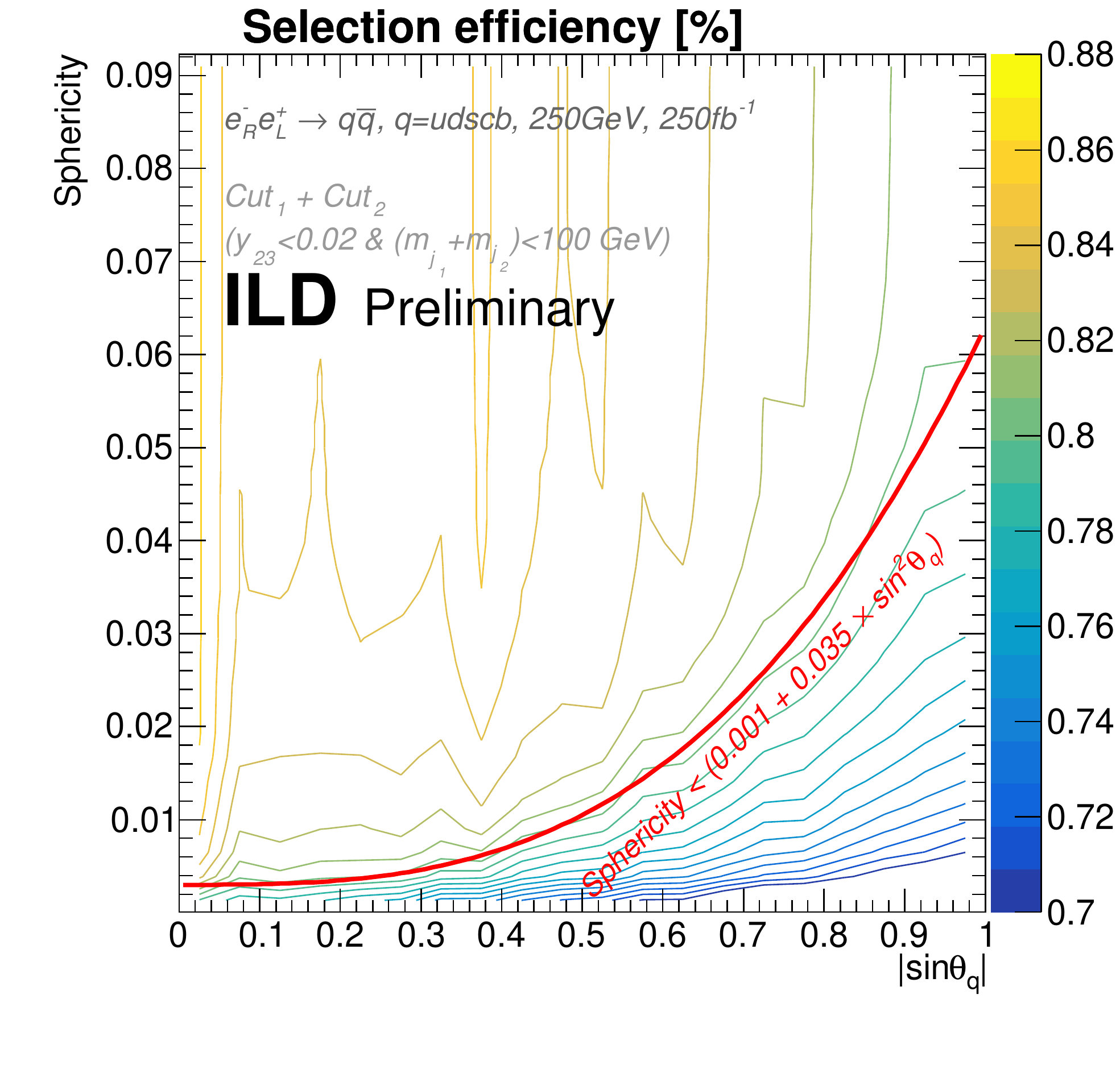}
     \end{tabular}
  \caption{\label{fig:spher} Efficiency of selection as a function of the $\theta$ angle and the sphericity.}
\end{figure}

A last cut in the sphericity of the event follows. The sphericity tensor is defined as
\begin{equation}
  S^{\alpha,\beta}=\frac{\Sigma_{i}p_{i}^{\alpha}p_{i}^{\beta}}{\Sigma_{i} |\vec{p_{i}}|^2} \,\,\,\,\,\, \alpha,\beta=1,2,3
\end{equation}
where $p_{i}^{\alpha}$ is the $\alpha$-component of the momentum of the $i$-particle or jet. The eigenvalues of the sphericity tensor are called $\lambda_{1}$, $\lambda_{2}$ and $\lambda_{3}$ (with $\lambda_{1}\geq\lambda_{2}\geq\lambda_{3}$ and $\lambda_{1}+\lambda_{2}+\lambda_{3}=1$). The sphericity is defined as $3/2(\lambda_{2}+\lambda_{3}$. Two jet like event have sphericity equal zero while completely isotropic events tend to have sphericity values equal to one.
The sphericity distribution for the different signals and backgrounds is shown in Figure \ref{fig:selection3}, last plot.
The presence of ISR radiation has an impact in the sphericity of the event by unbalancing
the momentum of the two jets. The ISR impact on the sphericity is dependent on the localisation of the event in the detector: two jets in the barrel
in an event with ISR will have larger sphericity than an event with the same amount of ISR but with
the two jets located in the forward/backward region. Therefore, applying a simple cut in the sphericity will
give a difference on acceptance between the central and forward regions. We, therefore, apply a differential cut.
The parametrisation of such cut is derived from Figure \ref{fig:spher} (top left plot).
The final efficiency as a function of \costheta after the differential cut is shown in Figure \ref{fig:selection}.
Similar cuts could be done using other related event shapes variables as the thrust or acolinearity. 
These options have been investigated and both perform similarly to the sphericity. 

\end{document}